\documentclass[a4paper,12pt]{article}
\usepackage{authblk}
\usepackage[square,sort,comma,numbers]{natbib}
\usepackage{soul}
\usepackage{color}
\usepackage{pifont}
\usepackage{mathptmx}
\usepackage{graphicx}
\usepackage{amsmath, amsfonts, amssymb, mathrsfs}
\usepackage{ccaption}
\usepackage{fixmath}
\usepackage{tabu}
\usepackage{longtable}

\usepackage{graphicx}
\DeclareMathOperator{\logit}{logit}
\usepackage{multirow}
\usepackage{multicol}
\usepackage[figuresright]{rotating}
\usepackage[framemethod=tikz]{mdframed}

\title{Efficient and Robust Approaches for Analysis of SMARTs: Illustration using the ADAPT-R Trial}
\author[1]{Lina M. Montoya \thanks{Email address for correspondence: lmontoya@unc.edu}}
\author[1]{Michael R. Kosorok}
\author[2]{Elvin H. Geng}
\author[3]{Joshua Schwab}
\author[4]{Thomas A. Odeny}
\author[3]{Maya L. Petersen}
\affil[1]{Department of Biostatistics, University of North Carolina at Chapel Hill, U.S.A.}
\affil[2]{Division of Infectious Diseases, Washington University in St. Louis, U.S.A.}
\affil[3]{Division of Epidemiology and Biostatistics, University of California, Berkeley, U.S.A.}
\affil[4]{National Cancer Institute, National Institutes of Health, U.S.A.}

\date{October 2022}

\begin{document}

\maketitle
\pagebreak

\begin{abstract}
Personalized intervention strategies, in particular those that modify treatment based on a participant's own response, are a core component of precision medicine approaches. Sequential Multiple Assignment Randomized Trials (SMARTs) are growing in popularity and are specifically designed to facilitate the evaluation of sequential adaptive strategies, in particular those embedded within the SMART. Advances in efficient estimation approaches that are able to incorporate machine learning while retaining valid inference can allow for more precise estimates of the effectiveness of these embedded regimes. However, to the best of our knowledge, such approaches have not yet been applied as the primary analysis in SMART trials. In this paper, we present a robust and efficient approach using Targeted Maximum Likelihood Estimation (TMLE) for estimating and contrasting expected outcomes under the dynamic regimes embedded in a SMART, together with generating simultaneous confidence intervals for the resulting estimates. We contrast this method with two alternatives (G-computation and Inverse Probability Weighting estimators). The precision gains and robust inference achievable through the use of TMLE to evaluate the effects of embedded regimes are illustrated using both outcome-blind simulations and a real data analysis from the Adaptive Strategies for Preventing and Treating Lapses of Retention in HIV Care (ADAPT-R) trial (NCT02338739), a SMART with a primary aim of identifying strategies to improve retention in HIV care among people living with HIV in sub-Saharan Africa.
\end{abstract}

%


\maketitle


%
\section{Introduction}
\label{s:intro}

One question central to precision medicine and public health asks: ``who should get which intervention, and in what sequence?" For example, a wide class of sequenced strategies start with an initial intervention, and then switch to a new, often higher intensity intervention based on participant response. These strategies are personalized because both the decision to switch interventions and the timing of the switch depend on an individual's own response. Data generated from a Sequential Multiple Assignment Randomized Trial (SMART) provide a straightforward way of evaluating the causal effects of such sequenced adaptive strategies (or dynamic regimes). Often, participants are given treatment (either randomly or deterministically) at pre-specified decision points based on their measured information (e.g., past treatments and/or intermediate covariates) up to that point. Assigning treatment sequentially based on a participant’s measured past -- including commonly, a patient's own response to earlier treatment -- defines a SMART’s embedded dynamic treatment regimes (or simply, embedded regimes; also known as adaptive interventions or strategies). These embedded regimes correspond to adaptive personalized strategies for assigning treatment, thus contributing to the goals of precision health. Critically, by design, SMARTs allow the effects of these embedded regimes (and others, such as optimal dynamic treatment regimes based on covariates beyond those that define the trial design; \cite{kosorok2019}) to be identified and estimated without risk of bias.

SMART designs are increasingly growing in popularity. For example, a recent review by \cite{bigirumurame2021sequential} cites 24 SMART protocol papers published since 2014. While primary analyses for SMARTs sometimes aim to examine the single timepoint static effects of the treatment options in the SMART's nested trials, they increasingly (in either primary or secondary aims) aim to evaluate the effects of embedded regimes (e.g., \cite{kasari2014communication, karp2019improving}) or additionally tailored individual interventions (e.g., \cite{sherwood2016bestfit}). When evaluating the SMART's embedded regimes, common approaches for estimating the expected counterfactual outcome (or ``value") of a given embedded regime use inverse probability weighting (IPW) estimators, including weighting and replicating approaches (introduced in \cite{robins2002analytic, van2007causal, bembom2007statistical}; see also \cite{nahum2012experimental}) and G-computation approaches (introduced in \cite{robins1986new, robins1987addendum, lavori2000design, lavori2004dynamic}). IPW estimators, and some G-computation estimators (depending on how the sequential regressions are estimated) will generally provide unbiased estimates of the value of the embedded regime; however, they are inefficient in that they do not make full use of baseline and time-updated covariates to improve estimator precision. Advances in semiparametric efficient substitution estimators, such as longitudinal targeted maximum likelihood estimation (TMLE), allow for the integration of machine learning in the estimation process, enabling more precise estimates while retaining valid inference (see \cite{petersen2015chapter} for a review in the context of SMARTs). Recent work has documented the potential of flexible covariate adjustment using machine learning, and TMLE in particular, to improve precision in single timepoint individually randomized trials (eg, \cite{benkeser2021improving}) and cluster randomized trials (eg, \cite{balzertwostage}). Simulations used to inform the design of SMARTs (see, e.g., \cite{petersen2015chapter, benkeser2020design}) further support the potential benefits of longitudinal TMLE for the primary analysis of embedded regimes in SMART studies. However, to the best of our knowledge, neither longitudinal TMLE nor other semiparametric efficient estimators have been implemented or reported as the primary analysis method of a published SMART.

In this paper we first review, using the ``Causal Roadmap" \citep{petersen2014causal}, how SMART designs can be used to identify the effects of embedded regimes, including the expected counterfactual outcome (or value) of each regime had all participants in the population followed it. We then describe an efficient and robust approach to estimating these counterfactual quantities without reliance on model assumptions, beyond the assumption of sequential randomization known by design. Specifically, we describe a longitudinal TMLE \citep{van2012targeted, bangrobins2005} for estimating the values of these embedded regimes. TMLE is a double robust, semi-parametric, efficient, plug-in estimator that incorporates machine learning to improve efficiency without sacrificing reliable inference. We review the assumptions needed for valid statistical inference using this estimator, and we show how to construct individual and simultaneous confidence intervals to evaluate multiple embedded regimes within a SMART. Specifically, we illustrate the use of longitudinal TMLE as the primary pre-specified analysis in the recently completed Adaptive Strategies for Preventing and Treating Lapses of Retention in HIV Care (ADAPT-R) trial (NCT02338739). We provide simulations to demonstrate the robustness of the approach, including an illustration of how outcome-blind simulations based on real trial data can be used to inform key decisions that must be pre-specified in a trial's analysis plan, such as specification of the machine learning methods empoyed for nuisance parameter estimation. We further provide a comparison to the commonly used IPW estimator. Using both simulations and analysis of the trial data, we illustrate how the pre-specified use of TMLE integrating machine learning, in the analysis of the ADAPT-R trial, resulted in substantial improvements in efficiency, and thereby trial power, and discuss the interpretation of trial results.

The article is organized as follows: in Section 2, we provide  background on the ADAPT-R trial. In Section 3, we describe the causal model, define the causal parameters corresponding to the value of each embedded regime, and identify statistical parameters. In Section 4, we discuss estimation and inference of the
identified statistical parameters. In Section 5, we present two simulation studies, with the dual objectives of illustrating the performance of these estimators and demonstrating how outcome blind simulations can be used to fully pre-specify a machine learning-based primary trial analysis using TMLE. In Section 6, we apply these methods to the ADAPT-R study. We close with a discussion.

\section{The ADAPT-R Trial}

The ADAPT-R trial was a SMART carried out to evaluate individualized sequenced behavioral interventions to optimize successful HIV care outcomes in Kenya. Up to 30\% of persons receiving HIV care in this population experience at least one lapse in HIV care; these lapses in retention can result in loss of viral suppression. Importantly, patients that experience a retention lapse have a diversity of characteristics and needs \citep{gengestimation}. As a result, there is no “one-size-fits-all” incentive or strategy to help patients stay in care and achieve virologic suppression, demonstrating the need for effective personalized treatment regimes to increase successful HIV care outcomes.

In ADAPT-R, 1,809 persons living with HIV and initiating antiretroviral treatment (ART) in the Nyanza region of Kenya were randomized to one of three initial interventions to prevent a lapse in care (short message service [SMS] text messages, conditional cash transfers [CCTs] in the form of transportation vouchers for on-time visits, or standard of care [SOC] education and counseling). Patients who had a lapse in care within the first year of follow-up were re-randomized to a more intensive intervention to facilitate return to care (SMS text messages paired with CCTs, peer navigation, or SOC outreach); patients who did not have a lapse in care during the first year and who received SMS or CCTs in the first randomization were re-randomized to either continue or discontinue that intervention (study design shown in Figure 1). 

Thus, in ADAPT-R there were 15 embedded regimes (see Table \ref{table0} for the complete list) that would have initially administered either SMS, CCTs, or SOC to all patients starting ART, and then either a) SMS with CCTs, peer navigators, or SOC in the second stage should a lapse occur, or b) for those on active first line treatment, a decision to continue or discontinue first stage treatment should no lapse occur. This article describes how to estimate the counterfactual probability of having suppressed viral replication (plasma HIV RNA level $<$ 500 copies/ml) two years after initial randomization, if a given embedded regime had been used for the full study population.

\begin{figure}[h]
    \centering
    \includegraphics[scale = .47]{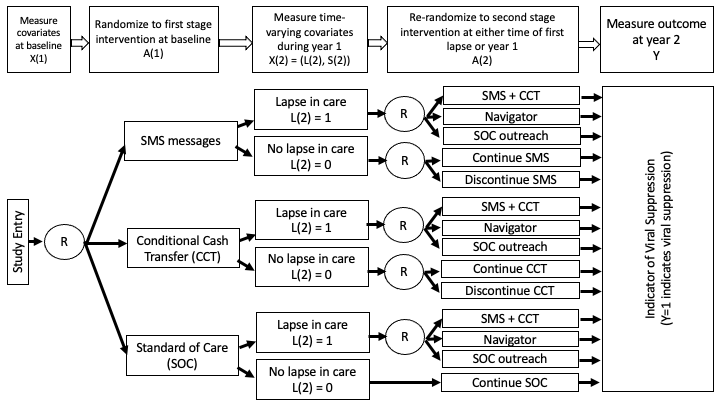}
    \caption{The Adaptive Strategies for Preventing and Treating Lapses of Retention in HIV Care (ADAPT-R) study design, a Sequential Multiple Assignment Randomized Trial (SMART). The circles with an ``R" denote points of randomization.}
    \label{adapt_fig}
\end{figure}

\begin{center}
\begin{table}[]{
\begin{tabular}{|l|l|l|l|}
\hline
\textbf{Embedded Regime ($\tilde{d}$) } & \textbf{Stage 1} & \textbf{Stage 2  if Lapse} & \textbf{Stage 2 if No Lapse} \\ \hline
1 & SOC & SOC outreach & Continue \\ \hline
2 & SMS & SOC outreach & Continue \\ \hline
3 & CCT & SOC outreach & Continue \\ \hline
4 & SOC & SMS + CCT & Continue \\ \hline
5 & SMS & SMS + CCT & Continue \\ \hline
6 & CCT & SMS + CCT & Continue \\ \hline
7 & SOC & Navigator & Continue \\ \hline
8 & SMS & Navigator & Continue \\ \hline
9 & CCT & Navigator & Continue \\ \hline
10 & SMS & SOC outreach & Discontinue \\ \hline
11 & CCT & SOC outreach & Discontinue \\ \hline
12 & SMS & SMS + CCT & Discontinue \\ \hline
13 & CCT & SMS + CCT & Discontinue \\ \hline
14 & SMS & Navigator & Discontinue \\ \hline
15 & CCT & Navigator & Discontinue \\ \hline
\end{tabular}
}
\caption{List of 15 dynamic treatment regimes embedded within the Adaptive Strategies for Preventing and Treating Lapses of Retention in HIV Care (ADAPT-R) study (i.e., ADAPT-R's 15 embedded regimes).}
\label{table0}
\end{table}
\end{center}

\section{Causal Roadmap}
\subsection{Causal Models}

The following structural causal model (SCM, denoted $\mathcal{M}^F$) will be used to describe the longitudinal process that gives rise to variables that are observed (endogenous) and not observed (exogenous) \citep{pearl2000}. The random variables in $\mathcal{M}^F$ follow the joint distribution $P_{U,X}$; the SCM describes the set of possible distributions for $P_{U,X}$. For a time $t$, the endogenous variables are 1) categorical interventions $A(t) \in \mathcal{A}_t$ (which could include right-censoring); 2) covariates $X(t) \in \mathcal{X}_t$, which include baseline covariates and time-varying covariates between interventions at time $t-1$ and $t$ (which could include indicators of time-dependent processes, such as death), and; 3) an outcome $Y \in \mathbb{R}$. Overbars are used to denote a variable's past history, e.g., $\bar{A}(t) = (A(1), \ldots, A(t))$ and $\bar{X}(t) = (X(1), \ldots, X(t))$, and $A(0) = X(0) = \emptyset$. Let $\bar{Z}(t) \subseteq (\bar{A}(t-1), \bar{X}(t))$ denote the subset of endogenous tailoring variables used by design in the SMART to assign treatment at time $t$.  Then, for observation time $t = 1, \ldots, K$, the following structural equations can describe a SMART's longitudinal data generating process:
\begin{align*}
    X(t) &= f_{X(t)}(U_{X(t)}, \bar{X}(t-1), \bar{A}(t-1)) \\
    A(t) &= f_{A(t)}(U_{A(t)}, \bar{Z}(t)) \\
    Y &= f_{Y}(U_{Y}, \bar{X}(K), \bar{A}(K)),
\end{align*}
where exogenous variables are denoted $U = (U_{X(t)}, U_{A(t)}, U_Y)$ and represent the unmeasured random input to the data generating system. Importantly, $f_{A(t)}$ represent known parametric functions (specifically, the randomization scheme used in the SMART). Further, $U_{A(t)}$ is known by design in a SMART to be independent of all other exogenous factors. We note that in a SMART, $t$ need not be the observation time; here, $t$ is the time of treatment assignment. Thus, in a $K$-stage SMART (without intervening on non-randomized intervention nodes, such as censoring), $K$ corresponds to the number of randomization stages.

\subsubsection{Data and Models - Application to ADAPT-R Study}\label{adaptdata} 
The ADAPT-R study provides an illustration of a SMART, where $t = 1$ is time of first randomization and $t = 2$ is time of second randomization (either date of first retention lapse or one year after initial randomization, whichever occurs first). In ADAPT-R, baseline covariates $X(1)$ included participant sex, age, WHO disease stage, CD4+ T cell count, an alcohol consumption measure, pregnancy status, and clinic site. The Stage 1 prevention intervention $A(1)$ consisted of either SMS text messages, CCTs, or SOC, each assigned with equal probability. Covariates assessed between randomization to Stage 1 and Stage 2 interventions, $X(2) = (L(2), S(2))$, included $L(2)$, an indicator of whether there was a lapse in care ($\geq$ 14 days late to a clinic visit) within the first year after enrollment, and $S(2)$, which included death, transfer to another clinic, time from first randomization to second randomization, pregnancy status at second randomization, plasma HIV RNA level at second randomization, and whether a participant could be successfully contacted prior to randomization. The Stage 2 retention intervention $A(2)$ consisted of either a) SOC outreach, SMS and CCT combined (SMS+CCT), or a peer navigator (Nav), each assigned with equal probability if there was a lapse in care ($L(2) = 1$); or, b) continuing or discontinuing the Stage 1 intervention, each assigned with equal probability, if there was no lapse in care ($L(2)=0$) and the initial intervention was either SMS or CCT; or, c) continuing SOC if there was no lapse in care and the initial intervention was SOC. The outcome of interest $Y$ was an indicator of remaining alive and with viral suppression at year 2.

The SCM for ADAPT-R can be written as follows, where $\bar{Z}(2) = (A(1), L(2))$:
\begin{align*}
    X(1) &= f_{X(1)}(U_{X(1)}) \\
    A(1) &= f_{A(1)}(U_{A(1)}, Z(1))  \\
    X(2) &= f_{X(2)}(U_{X(2)}, X(1), A(1)) \\
    A(2) &=  f_{A(2)}(U_{A(2)}, \bar{Z}(2))\\
    Y &= f_Y(U_Y, \bar{X}(2), \bar{A}(2)),
\end{align*}
where $A(1)$ is drawn from a Multinomial distribution with $n=1$, $k=3$, $p_{SMS}=p_{CCT}=p_{SOC}=1/3$ and $A(2)$ is drawn from a Bernoulli distribution with $p_{discont.}=p_{cont.} = 0.5$ if $L(2)=0$ and $A(1) \in \{$SMS, CCT$\}$, deterministically equal to SOC if $L(2)=0$ and $A(1) =$ SOC, and drawn from a Multinomial distribution with $n=1$, $k=3$ and $p_{SMS+CCT}=p_{Nav.}=p_{SOC}=1/3$ if $L(2)=1$. Additionally, here $(U_{A(1)}, U_{A(2)})$ are independent of each other and all other $U$s.

\subsection{Causal Questions and Parameters}\label{causalqs}

The focus of this manuscript is on evaluating outcomes under the dynamic treatment regimes embedded in a SMART. In other words, our causal questions take the form: what are the expected outcomes at the end of follow up if all members of the target population had followed each of the dynamic regimes embedded in the SMART design? 

Let $\phi_t(\bar{A}(t-1), \bar{X}(t)) \subseteq \mathcal{A}_t$ denote the set of allowable treatments for a participant presenting with $(\bar{A}(t-1), \bar{X}(t)) = (\bar{a}(t-1), \bar{x}(t))$ at time $t$. Then, let a decision rule $d_t$ be a function that takes as input the information accrued on a participant up to time $t$ and outputs a single treatment level from among the set of possible treatment levels to which a participant could be randomly assigned, for all covariate and treatment histories, i.e. $d_t:(\mathcal{X}_1 \times \ldots \times \mathcal{X}_t, \mathcal{A}_1 \times \ldots \times  \mathcal{A}_{t-1}) \rightarrow \phi_t(\bar{A}(t-1), \bar{X}(t))$. Denote $\mathcal{D}_t$ as the set of all such decision rules at time $t$. Let $d = (d_1, d_2, ..., d_K)$ be a dynamic treatment regime (i.e., a sequence of rules for assigning a treatment level at each randomization stage), and let $\bar{d}_t: (\mathcal{X}_1 \times \ldots \times \mathcal{X}_t, \mathcal{A}_1 \times \ldots \times  \mathcal{A}_{t-1}) \rightarrow \phi_1(X(1)) \times \ldots \times \phi_t(\bar{A}(t-1), \bar{X}(t))$ denote a regime sequence until time $t$. Let $\mathcal{D}$ be the set of all such dynamic treatment regimes.

We focus here on the \emph{embedded} dynamic treatment regimes in a SMART, which are particular sequences of rules with input $\bar{Z}(t)$, the tailoring variables used for assigning treatment at time $t$ in the actual randomization scheme employed by the trial, and output in $\phi_t(\bar{A}(t-1), \bar{X}(t))$, for all $t$. The set of embedded regimes $\tilde{\mathcal{D}}$ are subset of the entire set of dynamic regimes, and we denote $\tilde{\bar{d}}_t(\bar{Z}(t))$ as an element of $\tilde{\mathcal{D}}$, where $\tilde{\bar{d}}_t(\bar{Z}(t)) = \{\tilde{d}_1(Z(1)), \tilde{d}_2(\bar{Z}(2)), ..., \tilde{d}_t(\bar{Z}(t))\}$ is a SMART's embedded regime until time $t$.

A counterfactual outcome under an embedded dynamic treatment regime $\tilde{d}$ is an individual's outcome if, possibly contrary to fact, the individual had been assigned treatment according to the embedded regime $\tilde{d}$. This counterfactual outcome,  denoted $Y_{\tilde{d}}$, can be derived under an intervention on the above SCM, in which at each randomization stage in the SMART, the randomized treatment assignment mechanism used in the SMART is replaced with a deterministic assignment of a single treatment level based on observed history; i.e., for $t = 1, ..., K$:
\begin{align*}
    X(t) &= f_{X(t)}(U_{X(t)}, \bar{X}(t-1), \bar{A}(t-1)) \\
    A(t) &= \tilde{d}_t(\bar{Z}(t)) \\
    Y_{\tilde{d}} &= f_{Y}(U_{Y}, \bar{X}(K).
\end{align*}
The target causal parameters that answer our aforementioned causal queries are summary measures of the post-intervention distribution contained within the SCM. Here, the relevant causal parameters are the expected counterfactual outcomes had all participants received each of the SMART's embedded dynamic regimes; i.e., for one $\tilde{d} \in \tilde{\mathcal{D}}$:
\begin{align}
    \Psi_{\tilde{d}}^F(P_{U,X}) = \mathbb{E}_{P_{U,X}}[Y_{\tilde{d}}],
    \label{psi}
\end{align}
and the vector of the counterfactual values of the $D$ embedded regimes is denoted $\Psi^F(P_{U,X}) = \{\Psi_{\tilde{d}^{(1)}}^F(P_{U,X}), ..., \Psi_{\tilde{d}^{(D)}}^F(P_{U,X})\}$.

Of note, although in the current manuscript we focus on evaluating the particular regimes embedded within a SMART, we are not limited to asking the above causal questions when analyzing a SMART; by design, SMARTs easily allow for answering many causal questions corresponding to alternative aims of the study, such as:
\begin{enumerate}
    \item Point treatment static regimes for the embedded nested trials. For example, in ADAPT-R: what is the counterfactual probability of either experiencing a lapse in retention by one year or viral non-suppression at one year (an interim outcome not affected by the second line intervention assignment) had everyone received each of the initial interventions (SMS, CCT, and SOC)?
    \item Point treatment optimal dynamic treatment rule. For example, in ADAPT-R:  what is the optimal way to assign initial SMS, CCT, or SOC to participants based on their measured baseline characteristics to minimize the probability of a retention lapse by year one or viral non-suppression at year one? 
    \item Longitudinal optimal dynamic treatment regime. For example, in ADAPT-R: what is the optimal way to assign Stage 1 and Stage 2 treatments, in sequence, based on the observed baseline and time-varying covariates to minimize viral suppression at year two?
\end{enumerate}
We refer the reader to \cite{kosorok2019} for an overview of possible methods for answering these questions, particularly those that estimate optimal dynamic treatment rules.

Further, it could also be of interest to contrast pairs of embedded regimes; for example, for a pair of embedded regimes numbered $i,j \in \{1,\ldots,D\}$, $i\neq j$, one possible causal parameter that contrasts the efficacy between the two strategies is $\mathbb{E}_{P_{U,X}}[Y_{\tilde{d}^{(i)}} - Y_{\tilde{d}^{(j)}}]$.  Such contrasts follow naturally from the approach described in the paper to estimate the regime-specific mean outcomes. These contrasts could be specified \textit{a priori}, or omnibus tests could be employed, such as comparing the best embedded regime versus the worst (without knowing in advance which is which) or whether there are any significant differences in any of the regime values.

\subsubsection{Causal Parameters - Application to ADAPT-R Study}

Within ADAPT-R, the set of allowable treatments at each timepoint given past participant information are as follows: $\phi_1(X(1)) = \{\text{SMS, CCT, SOC}\}$, $\phi_2(X(1),A(1),S(2),L(2) = 1) = \{\text{SMS+CCT, Nav., SOC}\}$, $\phi_2(X(1),A(1)\in \{\text{SMS, CCT}\},S(2),L(2) = 0) = \{\text{Continue, Discontinue}\}$, and $\phi_2(X(1),A(1)= \{\text{SOC}\},S(2),L(2) = 0) = \{\text{Continue}\}$. 

Then, let $d = (d_1, d_2)$ be a dynamic treatment regime that uses participant information to assign the allowed treatments $\phi_1(X(1))$ and $\phi_2(A(1),\bar{X}(2))$ at Stage 1 and Stage 2, respectively; i.e., $d$ assigns $A(1)$ and $A(2)$ based on baseline covariates $X(1)$ and time varying covariates and initial treatment $\{\bar{X}(2), A(1)\}$, respectively. 

Specifically, we are interested in evaluating the particular sequence of rules that were used for assigning treatment within the SMART, $\tilde{d} \in \tilde{\mathcal{D}}$, i.e., the embedded regimes within ADAPT-R. The decision rules within the embedded regimes are characterized as follows: 1) at time 1, treat with either SMS, CCT, or SOC, regardless of baseline covariates, i.e., $\tilde{d}_1: Z(1) \rightarrow \phi_1(X(1))$, where $Z(1) = \emptyset$; and, 2) at time 2, treat with either SMS+CCT, Nav., SOC outreach, Continue, or Discontinue, depending on the initial treatment decision and whether there is a lapse in care in year 1, i.e., $\tilde{d}_2: Z(2) \rightarrow \phi_2(\bar{X}(2), A(1))$, where $Z(2) = \{A(1),L(2)\}$. For example, one embedded dynamic treatment regime $\tilde{d}$ assigns treatment via the following strategy: 1) $\tilde{d}_1 = $ assign SMS to everyone; 2) $\tilde{d}_2 = $ assign SMS+CCT if $L(2) = 1$ (lapse in care), continue SMS otherwise (succeed in care). 

For one embedded regime, $\Psi_{\tilde{d}}^F$ answers the causal question: what is the probability of viral suppression at two years of follow-up had everyone been assigned the same Stage 1 intervention, then each person assigned a Stage 2 intervention based on the participant's Stage 1 intervention and whether or not that person had a lapse in care? Further, we are interested in the vector $\Psi^F$, which contains the counterfactual probabilities of 2 year viral suppression had everyone received each of the 15 strategies listed in Table \ref{table0}.

Finally, one may be interested in contrasting two adaptive strategies for preventing lapses in HIV care. For example, to compare the efficacy of the first two regimes in Table \ref{table0}, let $\tilde{d}^{(1)}$ be regime \#1 from Table \ref{table0} (SOC, then SOC outreach if lapse and Continue if no lapse), and $\tilde{d}^{(2)}$ be regime \#2 (SMS, then SOC outreach if lapse and Continue if no lapse). Then the causal parameter that contrasts these two strategies is $\mathbb{E}_{P_{U,X}}[Y_{\tilde{d}^{(1)}} - Y_{\tilde{d}^{(2)}}]$.

\subsection{Statistical Model, Identification, and Statistical Target Parameter}\label{identification}

We assume the observed data $O_i \equiv (\bar{X}(K)_i, \bar{A}(K)_i, Y_i) \sim P_0 \in \mathcal{M}$, $i = 1, \ldots, n$ were generated by sampling $n$ independent and identically distributed copies from a data-generating system contained in $\mathcal{M}^F$ above. Here, $P_0$ is the observed data distribution, an element of $\mathcal{M}$, the statistical model.


Two conditions are necessary for identification; that is, for determining that the causal parameter (i.e., Equation \ref{psi}, a function of the counterfactual distribution, $P_{U,X}$) is equivalent to a statistical parameter (a function of the observed data distribution $P_{0}$) for all distributions $P_{U,X}$ contained in $\mathcal{M}^F$. For $t = 1,\ldots, K$ and $\tilde{d} \in \tilde{\mathcal{D}}$, we consider the 1) the sequential randomization assumption (SRA): $Y_{\tilde{d}_t} \perp A(t) |\bar{X}(t), \bar{A}(t-1) = \tilde{\bar{d}}_{t-1}(\bar{Z}(t-1))$; and, 2) sequential positivity assumption: $g_0(A(t)=\tilde{d}_t(\bar{Z}(t))|\bar{X}(t),\bar{A}(t-1) = \tilde{\bar{d}}_{t-1}(\bar{Z}(t-1))) > 0 - a.e.$, where $g_{A(t),0}(A(t)|\bar{X}(t),\bar{A}(t-1)) = P_0(A(t)|\bar{X}(t),\bar{A}(t-1))$ is the true conditional probability of the treatment at time $t$ given measured time-varying variables used in the study design. Informally, the SRA states that there are no unmeasured common causes between assignment of $A(t)$ and $Y$, given that the individual has followed the regime up to $t$ and information accrued up to $t$. The sequential positivity assumption states that among subjects who have followed the regime up to $t$, there must be a positive probability of continuing to follow that regime at $t$, regardless of a participant's past information. 

A SMART, by design, ensures that both conditions are met. For example, in ADAPT-R, individuals are completely randomized to $A(1)$ and are randomized based on measured, accrued information (i.e., $\bar{Z}(2)$) to $A(2)$. The probability of receiving any of the embedded decision rule treatments at Stage 1 given baseline covariates $X(1)$ is 1/3; the probability of receiving any of the possible embedded decision rule treatments at Stage 2 is 1/3 among people who had a lapse, 1/2 among people who succeeded in care and were initially given SMS or CCT, and 1 among people who succeeded in care and were initially given SOC. The general statistical parameter corresponding to $\Psi_{\tilde{d}}^F(P_{U,X})$ for one embedded regime $\tilde{d} = \tilde{\bar{d}}_K(\bar{Z}(K))$ is the G-computation formula \citep{robins1986new}:\begin{align}
\Psi_{\tilde{d}}(P_0)=& \sum_{x_1,\ldots, x_K}\mathbb{E}_{0}\left[Y|\bar{X}(K) = \bar{x}(K), \bar{A}(K) = \tilde{\bar{d}}_K(\bar{Z}(K))\right] \nonumber \\ 
    & \times \prod_{t=1}^K P_{0}\left(X(t) = x(t) | \bar{X}(t-1) = \bar{x}(t-1), \bar{A}(t-1) = \tilde{\bar{d}}_{t-1}(\bar{Z}(t-1))\right), 
    \label{gcomp}
\end{align} where the summation generalizes to an integral for continuous $X(t)$. Equation \ref{gcomp} can also be re-written as a series of iterated conditional expectations (ICEs; or sequential regressions) \citep{bangrobins2005}:
\begin{align}
    \Psi_{\tilde{d}}(P_0) = & \mathbb{E}_0[\mathbb{E}_0[ \ldots \nonumber\\
    & \mathbb{E}_0\left[\mathbb{E}_0\left[Y|\bar{X}(K), \bar{A}(K) = \tilde{\bar{d}}_K(\bar{Z}(K))\right] \vert \bar{X}(K-1), \bar{A}(K-1) = \tilde{\bar{d}}_{K-1}(\bar{Z}(K-1)) \right] \nonumber\\
    & \ldots \vert X(1), A(1) = \tilde{d}_1(Z(1)) ]],
     \label{ice}
\end{align}
or as the following IPW estimand: $\Psi_{\tilde{d}}(P_0) = \mathbb{E}_0\left[ \frac{\mathbb{I}[\bar{A}(K) = \tilde{\bar{d}}_K(\bar{Z}(K))]}{\prod_{t=1}^K g_0(A(t) \vert \bar{X}(t), \bar{A}(t-1))}Y\right]$. 

The observed data for an ADAPT-R participant are $O = \{\bar{X}(2), \bar{A}(2), Y\}$; the observed dataset consists of 1,809 i.i.d. observations of $O$ generated by a process described by the aforementioned causal model. The statistical target parameter corresponding to the value of (i.e., the expectation of the counterfactual outcome under) an embedded regime within ADAPT-R is: 
\begin{align*}
\Psi_{\tilde{d}}(P_0)= & \sum_{x_1,x_2}\mathbb{E}_{0}\left[Y|\bar{X}(2) = \bar{x}(2), \bar{A}(2) = \tilde{\bar{d}}_2(\bar{Z}(2))\right] \\ 
& \times P_{0}\left(X(2) = x(2) | x(1), A(1) = \tilde{d}_1(Z(1))\right)P_{0}\left(X(1) = x(1)\right).
\end{align*}
The vector of all embedded regime values is identified as $\Psi(P_0) = (\Psi_{\tilde{d}^{(1)}}(P_0), ..., \Psi_{\tilde{d}^{(D)}}(P_0))$; in ADAPT-R, $D = 15$. Finally, if one were interested in comparing the value of two embedded regimes $\tilde{d}^{(i)}$ and $\tilde{d}^{(j)}$, the statistical parameter corresponding to this contrast would be $\Psi_{\tilde{d}^{(i)}}(P_0) - \Psi_{\tilde{d}^{(j)}}(P_0)$.

\section{Estimation and Inference for the Values of Embedded Regimes}\label{estandinf}



We are interested in estimators for the statistical parameter identified in Section \ref{identification} -- i.e., estimators for evaluating a SMART's embedded dynamic regimes. We focus on a longitudinal TMLE, and compare this with the IPW and G-computation estimator based on ICEs. All of these estimators can be implemented with the \emph{ltmle} package \citep{ltmlepackage, petersen2014targeted}. We briefly describe the longitudinal TMLE employed in the ADAPT-R analysis here, and we refer the reader to the Appendix A for a detailed description of the steps for implementing the three estimators.

The longitudinal TMLE employed here is a flexible and robust approach that estimates the value of a sequential regime by fitting initial estimates of the series of ICEs (Equation \ref{ice}) and updating these using either the known or estimated treatment mechanisms \citep{bangrobins2005, van2012targeted}.  Critically, TMLE allows for the use of flexible machine learning methods, such as SuperLearner \citep{van2007super}, to generate the initial estimates of the ICEs. Once updated using the treatment mechanisms, these estimates are then used to implement a plug-in estimator of the target parameter, as defined in Equation \ref{ice}. In contrast, the G-computation estimators exclusively rely on initial (untargeted) estimates of the ICEs, while the IPW estimator relies on either the estimated or true (and known, in a SMART) treatment mechanisms to estimate the value of an embedded regime.

Inference for the TMLE estimates of the embedded regime values can be based on estimates of the efficient influence curve for the target statistical parameter \citep{bangrobins2005}, which can be used to construct Wald-type 95\% confidence intervals that, under assumptions, provide nominal to conservative coverage for the value of one embedded regime. Further, because one goal is to evaluate the multiple dynamic regimes embedded in a SMART at the same time, one can also use an estimate of the efficient influence curve to construct simultaneous confidence intervals \citep{cai2020one}. For example, in ADAPT-R, one goal might be to evaluate 15 embedded regimes simultaneously; simultaneous confidence intervals aim to ensure that all estimated confidence intervals contain the true values of the embedded regimes at the nominal level, thus providing one approach to account for multiplicity. The same approach can be easily extended to handle multiple comparisons of these regimes. We refer the reader to Appendix B for technical details on how to construct these confidence intervals.


If implemented carefully, longitudinal TMLE has the potential to substantially improve efficiency (both asymptotically and in finite samples), and thereby increase study power. In a SMART, the treatment mechanism is known; thus, if there is no censoring, one could use the true conditional treatment probabilities $g_0(A(t)|\bar{X}(t), \bar{A}(t-1)) \equiv g_0(A(t)|\bar{Z}(t))$ in either the IPW or TMLE estimators. Estimator precision can be improved, however, by estimating the treatment mechanism using a maximum likelihood estimate of the parameters of a correctly specified parametric model (such as a generalized linear model including either $Z(t)$ alone, or including additional covariates in $X(t)$) \citep{van2003unified}. Either of these estimator specifications (either usage of the true treatment mechanism probabilities or estimates via correctly specified parametric models) will result in IPW and TMLE estimators that are consistent; however, the use of TMLE allows for further efficiency gains through additional estimation of the ICEs. Informally, as long as: 1) either the ICE initial estimates are not overfit, or sample splitting is incorporated in the estimator such that the targeted update is fit on  data independent of that used in the initial fit \citep{zheng2010asymptotic}; and, 2) the ICEs are estimated consistently -- then the resulting TMLE will be efficient. Importantly, when using TMLE (or other double robust semiparametric efficient estimators), the iterated conditional expectations can be estimated using machine learning, increasing the chance that they are estimated consistently and potentially further improving finite sample variance.

We contrast these efficiency properties of TMLE with those of the G-computation and IPW estimators. In particular, there is no valid theory for inference on the G-computation estimator if the ICEs are estimated using either flexible machine learning algorithms or with misspecified parametric models. IPW estimators do allow for conservative or nominal inference with either consistently estimated or true values of the treatment mechanisms, but they are not efficient.

\section{Simulations}\label{simulations}

Using simulations, we evaluated the performance of various estimators for the values of a SMART's embedded regimes. We did this for two data generating processes (DGPs) corresponding to SMART designs (i.e., in which the true treatment mechanism $g_0(A(t)|\bar{Z}(t))$ is assumed known): 1) a simple, hypothetical DGP in which re-randomization is based only on intermediate covariates (and not initial treatment; DGP 1); and 2) an outcome-blind simulation based on data from ADAPT-R in which the actual ADAPT-R empirical covariate distribution and known treatment mechanism was used, but for which the outcomes themselves were simulated (DGP 2). Importantly, the latter represents a powerful tool for fine-tuning choices for pre-specification of an estimator.

For both data generating processes, we first implemented the IPW estimator using the true treatment mechanisms, the $g_0(A(t)|\bar{Z}(t))$ factors, which are known in a SMART (denoted ``Min. adj. IPW w/ $g_0$"). Second, we implemented an IPW estimator where $g_0(A(t)|\bar{Z}(t))$ was estimated using the empirical proportions of each covariate and treatment history strata (i.e., a saturated model, denoted ``Min. adj. IPW w/ $g_n$"). We note that the G-computation estimator and TMLE will generate equivalent estimates to this IPW estimator if the ICE factors are also estimated with saturated regression models. Third, we implemented an IPW estimator where $g_0(A(t)|\bar{Z}(t))$ was estimated using main-terms logistic regression models of treatment on all past treatments and covariates (including covariates not used in the randomization scheme, but predictive of the outcome). In subsequent results, this is denoted ``Full adj. IPW." Fourth, we implemented the G-computation estimator that adjusted for all covariates through ICE factors estimated with SuperLearner (see Appendix C for specifications including the library of algorithms used). Finally, we implemented a TMLE in which ICEs were estimated adjusting for all covariates using SuperLearner, and $g$ was estimated using a correctly specified logistic regression model noting that this parametric model specification was known to contain the true treatment mechanism. For TMLE, in DGP 1 all baseline and time-varying covariates were used in the estimation of the $g$ factors, while in DGP 2 estimates of $g$ used the minimal adjustment set (i.e., $\bar{Z}(2)$) to avoid overfitting.

We evaluated estimator performance in terms of absolute bias, variance, confidence interval width, and 95\% confidence interval coverage (for both individual and simultaneous confidence intervals). Inference was based on the influence-curve procedures described in Section \ref{estandinf}. If IPW was employed, the estimated IPW influence curve was used for inference; if TMLE was employed, the estimated efficient influence curve was used for inference. As noted in Section \ref{estandinf}, we do not provide inference results for the G-computation estimator.

We refer the reader to Appendix C for details on how the simulations were implemented, including specific DGPs and algorithm configurations. We used the \emph{ltmle} R package for estimation and inference \citep{ltmlepackage, petersen2014causal}. Each simulation consisted of 1,000 iterations of $n$=1,692 observations (the sample size for the ADAPT-R's analysis dataset after excluding 117 persons for a missing outcome measure). We additionally implemented these simulations for DGP 1 with a reduced sample size of $n=$750 and various different SuperLearner library configurations for comparison.

\subsection{Simulation Results}\label{simresults}


The results described below are shown in Figures \ref{DGP1fig} and \ref{DGP2fig}, tables in Appendix C, and additional figures in Appendix D. 

The untargeted G-computation estimator, in which the ICEs were fit using machine learning, exhibited the highest bias among the estimators across the embedded regimes evaluated. Specifically, for DGP 1, the hypothetical SMART with minimal covariates, the mean absolute difference between the G-computation estimate and the truth ranged from 0.09\% to 0.69\% (1.70-91.01 times that of the bias of any IPW or TMLE estimator). For DGP 2, which incorporated covariates re-sampled from the empirical distribution of ADAPT-R data, the absolute mean difference between the G-computation estimate and the truth ranged from 0.20\% to 1.41\%; across all embedded regimes except for 2 and 6, bias of the G-computation estimator was 1.62-138.64 times that of the bias of IPW or TMLE. For the remaining embedded regimes 2 and 6, bias of the G-computation estimator was minimal and similar to the IPW estimator and TMLE, likely because the SuperLearner consistently chose saturated regression models to estimate the ICEs, thus generating equivalent estimates to TMLE and IPW.

As expected, the IPW that used the known treatment mechanism (``Min. adj. IPW w/ $g_0$") was unbiased with close to nominal coverage (93.1\%-95.8\% range across both DGPs and both confidence interval types), but exhibited a higher variance than other estimators. For example, the relative variance of the IPW estimator that used the known, true probabilities of treatment versus the IPW estimator that used empirical proportions given the minimal adjustment set $\bar{Z}(2)$ to estimate the $g$ factors (i.e., ``Min. adj. IPW w/ $g_0$" versus ``Min. adj. IPW w/ $g_n$") was 2.10-6.14 for DGP 1 and 2.87-3.37 for DGP 2. Finally, the IPW estimator in which the $g$ factors were estimated adjusting for additional baseline and time-varying covariates (i.e., ``Full adj. IPW") resulted in some variance reduction compared to the IPW in which $g$ was estimated using only the minimal adjustment set $\bar{Z}(2)$ (i.e., the ``Min. adj. w/ $g_n$" variance was up to 1.14 that of the ``Full adj. IPW" variance). Although estimation of the treatment mechanism reduced the variance of the point estimator, reductions in the 95\% confidence interval widths (and by implication, power for contrasting regimes) were limited by the fact that influence curve-based inference for IPW estimators in which $g$ was estimated yielded conservative inference (i.e., 99.3\%-100.0\% confidence interval coverage across all DGPs and confidence interval types).

The TMLE with the estimated treatment mechanisms and ICEs estimated using machine learning (SuperLearner) were unbiased with close to nominal confidence interval coverage (93.4\%-96.0\% across both DGPs and both confidence interval types). In addition, TMLE showed variance gains relative to IPW, particularly for DGP 2 (e.g., the relative variance of the fully adjusted IPW versus TMLE was 1.01-1.12 for DGP 1 and 1.36-1.58 for DGP 2), through its ICE estimation using machine learning (i.e., beyond estimation of the $g$ factors, as in IPW estimation). TMLE resulted in substantially narrower mean 95\% confidence intervals across both DGPs (1.57-2.62 times that of fully adjusted IPW confidence interval widths), due both to a slightly lower variance of the estimator, and the less conservative influence curve-based variance estimation compared to IPW. 

In further simulations (see results in Appendix D), a reduction in sample size to $n = 750$ resulted in similar comparative performance across the estimators. Inclusion of a tree-based method in the SuperLearner library (namely, recursive partitioning and regression trees; \cite{breiman2017classification}) or highly adaptive lasso (HAL; \cite{benkeser2016highly}) yielded similar patterns;
though, notably, when including a tree-based method the bias increased significantly for the G-computation estimator.

\begin{figure}[h]
    \centering
    \includegraphics[scale = .65]{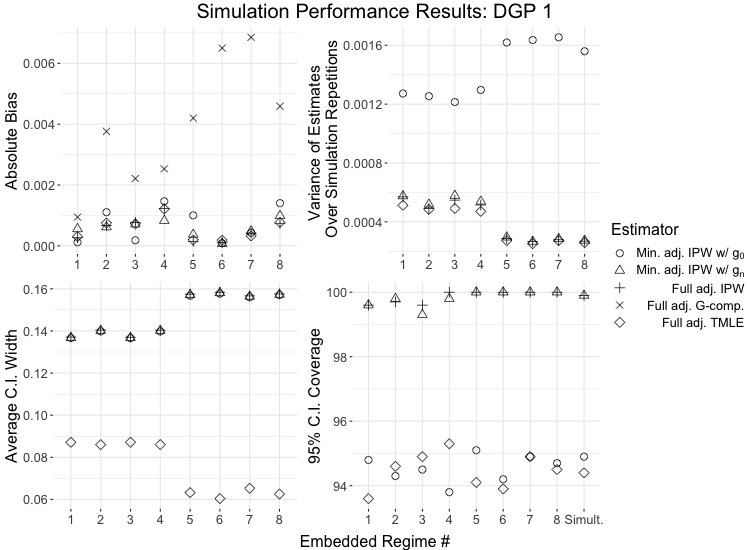}
    \caption{DGP 1. Performance (top left panel is absolute bias, top right panel is Monte Carlo variance (over simulation repetitions), bottom left panel is mean confidence interval [CI] width across simulation repetitions, and bottom right panel is 95\% CI coverage) of candidate estimators of the value of each of the 8 embedded regimes within the simple sequential multiple assignment randomized trial (SMART) generated from DGP 1. The 5 estimators evaluated are: 1) an inverse probability weighted (IPW) estimator with weights based on the true, known probability of receiving treatment given the initial treatment and lapse response (``Min. adj IPW w/ $g_0$"); 2) an IPW estimator with estimated weights based on the empirical proportion of receiving treatment given the initial treatment and lapse response, which is equivalent to a TMLE or G-computation estimator where iterated conditional expectation (ICE) factors are estimated with saturated regression models (``Min. adj IPW w/ $g_n$"); 3) an IPW estimator with estimated weights that adjust for all covariates (``Full adj. IPW"); 4) a G-computation estimator based on ICEs estimated with machine learning that adjust for all covariates (``Full adj. G-comp."); and 5) a targeted maximum likelihood estimator (TMLE) that adjusts for all covariates (``Full adj. TMLE"). Both individual and simultaneous CI coverage is shown under the regime numbers 1-8 and ``Simult.," respectively.}
    \label{DGP1fig}
\end{figure}

\begin{figure}[h]
    \centering
    \includegraphics[scale = .4]{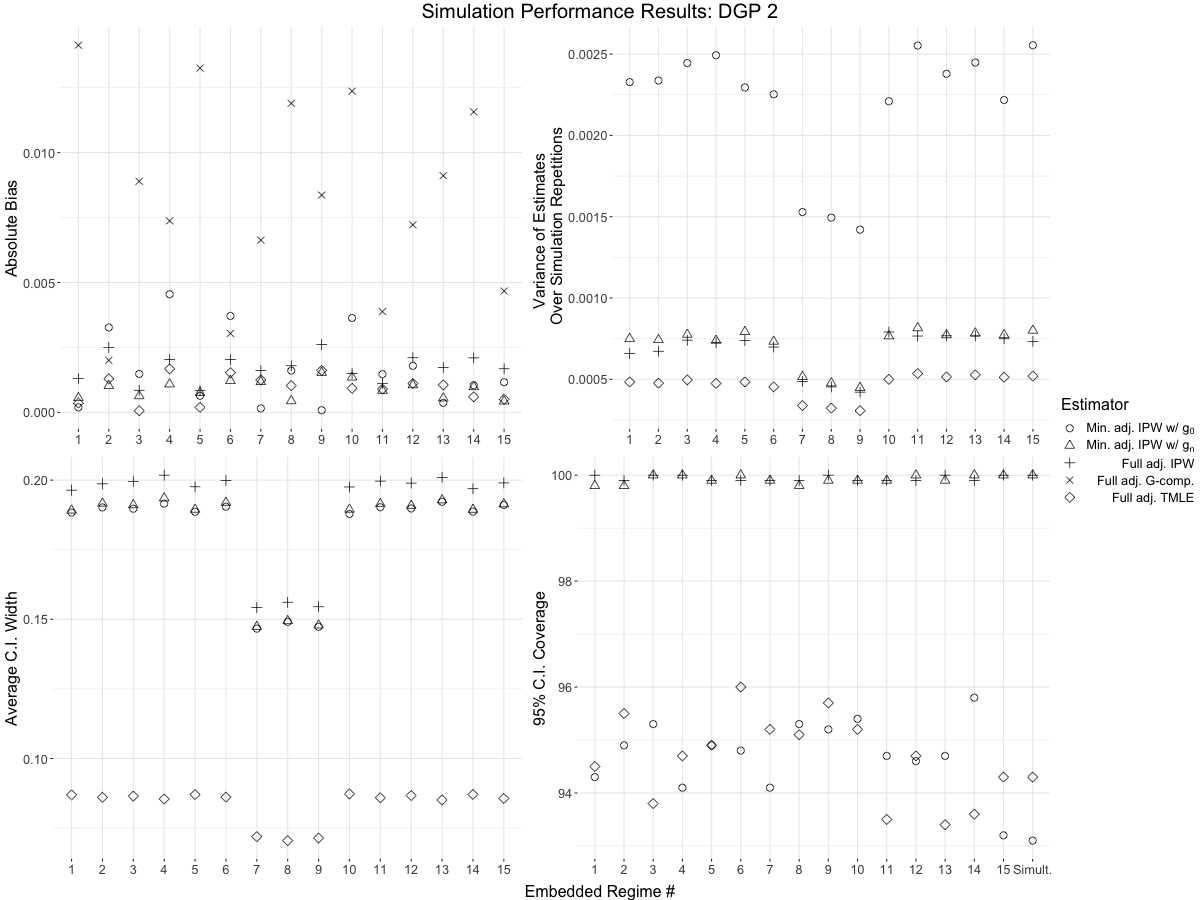}
    \caption{DGP 2. Performance (top left panel is absolute bias, top right panel is Monte Carlo variance (over simulation repetitions), bottom left panel is mean confidence interval [CI] width across simulation repetitions, and bottom right panel is 95\% CI coverage) of candidate estimators of the value of each of the 15 embedded regimes within the outcome-blind simulation of the Adaptive Strategies for Preventing and Treating Lapses of Retention in HIV Care (ADAPT-R) trial (DGP 2). The 5 estimators evaluated are: 1) an inverse probability weighted (IPW) estimator with weights based on the true, known probability of receiving treatment given the initial treatment and lapse response (``Min. adj IPW w/ $g_0$"); 2) an IPW estimator with estimated weights based on the empirical proportion of receiving treatment given the initial treatment and lapse response, which is equivalent to a TMLE or G-computation estimator where iterated conditional expectation (ICE) factors are estimated with saturated regression models (``Min. adj IPW w/ $g_n$"); 3) an IPW estimator with estimated weights that adjust for all covariates (``Full adj. IPW"); 4) a G-computation estimator based on ICEs estimated with machine learning that adjust for all covariates (``Full adj. G-comp."); and 5) a targeted maximum likelihood estimator (TMLE) that adjusts for all covariates (``Full adj. TMLE"). Both individual and simultaneous CI coverage is shown under the regime numbers 1-8 and ``Simult.," respectively.}
    \label{DGP2fig}
\end{figure}

\section{ADAPT-R Study Results}\label{adaptresults}

Of the 1,809 ADAPT-R participants, 117 did not have a viral load outcome; these patients were excluded from the analytic dataset ($n$=1,692; noting that if an outcome variable has substantial missingness, one could incorporate this into the causal model and thus adjust for it).  Using this sample, we conducted two analyses for the current manuscript, described below. A full report and interpretation of ADAPT-R's main results, including the clinical and public health implications, will be published in a separate manuscript.

\begin{figure}[h]
    \centering
    \includegraphics[scale = .60]{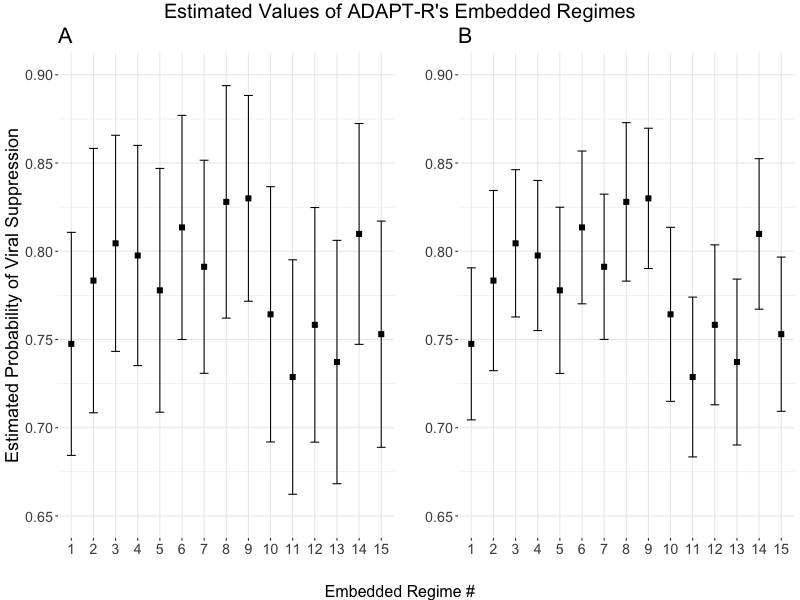}
    \caption{Analysis of the Adaptive Strategies for Preventing and Treating Lapses of Retention in HIV Care (ADAPT-R) study. Estimates of the probability of viral suppression under each of ADAPT-R's 15 embedded regimes listed in Table \ref{table0}. The squares are TMLE point estimates and the error bars are 95\% confidence intervals on these point estimates (simultaneous and single confidence intervals in Panels A and B, respectively. We note that these point estimates vary slightly from pre-specified analyses in ADAPT-R in that the latter used SuperLearner rather than logistic regressions for estimation of the treatment mechanism).}
    \label{plotADAPT}
\end{figure}

First, we estimated (using TMLE, as described for the ``Full adj. TMLE" estimator in simulations for DGP 2) and obtained inference on (using influence curve-based single and simultaneous confidence intervals) the value of each of ADAPT-R's 15 HIV care retention strategies (embedded regimes). The results of this analysis are shown in Figure \ref{plotADAPT} and Appendix E (Table 5, which includes the number of patients who contributed to each of the regimes). A point estimate reflects the estimated probability of viral suppression had the study population followed one of ADAPT-R's embedded regimes. For example, for the second embedded regime, we estimate that an intervention to deliver SMS to the full target population at time of ART initiation, followed by a transition from SMS to SOC outreach if a lapse in retention occurred, or the continuation of SMS if a lapse did not occur, would have resulted in 78.34\% (95\% simultaneous CI: [70.84\%, 85.83\%]) of the population alive and with a suppressed viral load after two years.

\begin{figure}[h]
    \centering
    \includegraphics[scale = .60]{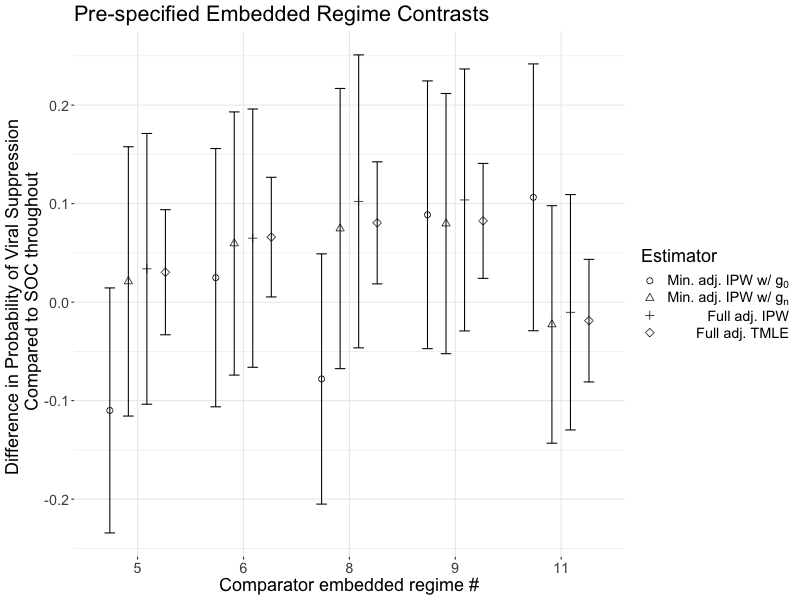}
    \caption{Pre-specified contrast analysis of the Adaptive Strategies for Preventing and Treating Lapses of Retention in HIV Care (ADAPT-R) study. Estimates of the difference in probability of viral suppression for the following pre-specified rules compared to standard of care (SOC) throughout: 1) short message service (SMS) with continuation if no lapse and addition of conditional cash transfer (CCT) if a lapse occurred (embedded regime number 5); 2) CCT with continuation if no lapse and addition of SMS if lapse occurred (embedded regime number 6); 3) SMS with continuation if no lapse and replacement with navigator if lapse occurred (regime number 8); 4) CCT with continuation of no lapse and replacement of navigator if lapse occurred (regime number 9); and 5) initial CCT, with SOC outreach if a lapse occurred and disconuation if no lapse occurred (regime number 11). Shapes are point estimates (and error-bars are influence curve-based individual confidence intervals), which were generated with: 1) an inverse probability weighted (IPW) estimator with weights based on the true, known probability of receiving treatment given the initial treatment and lapse response (``Min. adj IPW w/ $g_0$"); 2) an IPW estimator with estimated weights based on the empirical proportion of receiving treatment given the initial treatment and lapse response (``Min. adj IPW w/ $g_n$"); 3) an IPW estimator with estimated weights that adjust for all covariates (``Full adj. IPW"); and 4) a targeted maximum likelihood estimator (TMLE) that adjusts for all covariates (``Full adj. TMLE").}
    \label{plotADAPT_contrast}
\end{figure}

Second, we evaluated five pre-specified contrasts (that is, differences in values of pre-specified pairs of rules) between the regimes, shown in Figure \ref{plotADAPT_contrast} and Appendix E (Table 6). The first four regime pairs were chosen to compare the ``fully active" regime arms versus SOC throughout; a fifth pre-specified contrast evaluated the effectiveness of a strategy of time-limited CCT (initial CCT, with SOC outreach if a lapse occurred and discontinuation if no lapse occurred) versus SOC throughout. We implemented all three IPW estimators and the fully adjusted TMLE, described in the above simulations. Due to the small number of pre-specified hypotheses tested, multiplicity correction was not employed in these tests. TMLE estimates suggest that an active first line therapy (such as SMS or CCT) followed by a tailored peer-navigator was effective in improving viral suppression compared to the current HIV care standard throughout. In contrast, a single time limited CCTs did not result in significantly different viral suppression compared to receiving SOC throughout.

We note the important variance reduction in the TMLE that used the full adjustment set versus all other candidate estimators. In particular, confidence interval widths for non-TMLE estimates were wider (i.e., 1.92 to 2.40 times wider) than the full adjustment set. Critically, had we used an IPW estimator using either the known treatment mechanism or an estimated treatment mechanism -- a version of weighting and replicating -- instead of the presented TMLE, the results would not have detected any statistically significant contrasts.

\section{Discussion}
\label{s:discuss}

The purpose of this paper was to illustrate implementation of longitudinal targeted maximum likelihood estimation, integrating machine learning, for the evaluation of the embedded dynamic regimes in a SMART study. While previously described, to the best of our knowledge, this approach has not yet been applied in the primary published analysis of a SMART. We additionally illustrated how to obtain simultaneous confidence intervals on the values of multiple dynamic treatment regimes embedded in a SMART. In simulations and real data, we found substantial precision benefits from using this double robust, efficient estimator, especially in conjunction with adjustment for time-varying and baseline covariates using flexible machine learning approaches.

Specifically, in simulations, TMLE and IPW showed reduced bias compared to a G-computation estimator that utilized non-targeted machine learning-based iterated outcome regressions. Estimation of the treatment mechanism (compared to using the known, true probabilities of treatment), reduced the variance of the IPW estimator, with some further reduction in variance achieved through adjustment for covariates in addition to the minimal set needed for satisfying the SRA. However, the width of the IPW confidence intervals remained higher than that of TMLE using the corresponding adjustment set. Precision gains in the TMLE were driven both by reduced variance of the point estimator and by less conservative influence curved-based variances estimates. When estimating both its treatment mechanism (via a correctly specified, baseline and time-varying covariate adjusted logistic regression) and iterated outcome regressions (via SuperLearner, adjusting for all covariates) TMLE maintained close to nominal confidence interval coverage. This is analogous to the efficiency gains seen when adjusting for baseline covariates predictive of the outcome in a standard, single time-point randomized trials \citep{moore2009covariate}.

In addition, we showed how to evaluate the embedded regimes of ADAPT-R, a SMART carried out in Kenya to prevent lapses in HIV care. ADAPT-R's embedded regimes consisted of a menu of strategies that adapted to patients' responses to HIV care. The purpose of evaluating these regimes was to see the effect of each of these tailored strategies on viral suppression among this population. Using TMLE (with and without simultaneous confidence intervals), we obtained estimates of the probability of viral suppression for each of ADAPT's 15 embedded regimes. Further, with all IPW and TMLE estimators presented in this manuscript, we contrasted the efficacy between pre-specified strategies. Notably, results of these analyses illustrated the precision benefits associated with the longitudinal TMLE. In particular, had we used any of a set of common IPW estimators -- estimators that do not fully leverage covariate data, machine learning, and semiparametric efficiency theory -- we would not have been able to learn that active sequence strategies tailored to having a lapse in care improve viral suppression, compared to the current HIV care standard throughout. This emphasizes the importance of the described efficiency gains within our HIV care research, and also given the recent increase in ``small n SMARTs" (e.g., \cite{chao2020dynamic}).

The results of the analysis of ADAPT-R's embedded regimes present a menu of individualized strategies to help patients remain in HIV care. In particular, these results demonstrate the necessity for following HIV patients longitudinally in their treatment program, in order to escalate their treatment when needed, and shed light on effective escalation strategies. Critically, this kind of insight is not one we could have gleaned without a SMART. The current work aims to contribute to help uncover the potential of SMART designs so that -- with more precision and certainty -- we are more equipped to learn which treatments work better for whom, and when.


\section*{Acknowledgements}

Research reported in this publication was supported by NIAID awards R01AI074345, K24AI134413, and F31AI140962. The content is solely the responsibility of the authors and does not necessarily represent the official views of the NIH.  \vspace*{-8pt}

\section*{Supporting Information}

The code, simulations, and results for this manuscript can be found at:\\
https://github.com/lmmontoya/SMART-sims.\vspace*{-8pt}

\bibliographystyle{unsrt}
\bibliography{main}

\begin{thebibliography}{10}

\bibitem{kosorok2019}
M~R Kosorok and E~B Laber.
\newblock Precision medicine.
\newblock {\em Annual Review of Statistics and its Application}, 6:263--286,
  2019.

\bibitem{bigirumurame2021sequential}
Theophile Bigirumurame, Germaine Uwimpuhwe, and James Wason.
\newblock Sequential multiple assignment randomized trial studies should report
  all key components: a systematic review.
\newblock {\em Journal of Clinical Epidemiology}, 142:152--160, 2022.

\bibitem{kasari2014communication}
Connie Kasari, Ann Kaiser, Kelly Goods, Jennifer Nietfeld, Pamela Mathy,
  Rebecca Landa, et~al.
\newblock Communication interventions for minimally verbal children with
  autism: A sequential multiple assignment randomized trial.
\newblock {\em Journal of the American Academy of Child \& Adolescent
  Psychiatry}, 53(6):635--646, 2014.

\bibitem{karp2019improving}
Jordan~F Karp, Jun Zhang, Abdus~S Wahed, Stewart Anderson, Mary~Amanda Dew,
  G~Kelley Fitzgerald, et~al.
\newblock Improving patient reported outcomes and preventing depression and
  anxiety in older adults with knee osteoarthritis: results of a sequenced
  multiple assignment randomized trial (smart) study.
\newblock {\em The American Journal of Geriatric Psychiatry},
  27(10):1035--1045, 2019.

\bibitem{sherwood2016bestfit}
Nancy~E Sherwood, Meghan~L Butryn, Evan~M Forman, Daniel Almirall, Elisabeth~M
  Seburg, A~Lauren Crain, et~al.
\newblock The bestfit trial: A smart approach to developing individualized
  weight loss treatments.
\newblock {\em Contemporary clinical trials}, 47:209--216, 2016.

\bibitem{robins2002analytic}
James~M Robins.
\newblock Analytic methods for estimating hiv-treatment and cofactor effects.
\newblock In {\em Methodological issues in AIDS behavioral research}, pages
  213--288. Springer, 2002.

\bibitem{van2007causal}
Mark~J van~der Laan and Maya~L Petersen.
\newblock Causal effect models for realistic individualized treatment and
  intention to treat rules.
\newblock {\em The International Journal of Biostatistics}, 3(1), 2007.

\bibitem{bembom2007statistical}
Oliver Bembom and Mark~J van~der Laan.
\newblock Statistical methods for analyzing sequentially randomized trials.
\newblock {\em Journal of the National Cancer Institute}, 99(21):1577--1582,
  2007.

\bibitem{nahum2012experimental}
Inbal Nahum-Shani, Min Qian, Daniel Almirall, William~E Pelham, Beth Gnagy,
  Gregory~A Fabiano, James~G Waxmonsky, Jihnhee Yu, and Susan~A Murphy.
\newblock Experimental design and primary data analysis methods for comparing
  adaptive interventions.
\newblock {\em Psychological Methods}, 17(4):457, 2012.

\bibitem{robins1986new}
James~M Robins.
\newblock A new approach to causal inference in mortality studies with a
  sustained exposure period - application to control of the healthy worker
  survivor effect.
\newblock {\em Mathematical Modelling}, 7:9--12, 1986.

\bibitem{robins1987addendum}
James~M Robins.
\newblock Addendum to “a new approach to causal inference in mortality
  studies with a sustained exposure period—application to control of the
  healthy worker survivor effect”.
\newblock {\em Computers \& Mathematics with Applications}, 14(9-12):923--945,
  1987.

\bibitem{lavori2000design}
Philip~W Lavori and Ree Dawson.
\newblock A design for testing clinical strategies: biased adaptive
  within-subject randomization.
\newblock {\em Journal of the Royal Statistical Society: Series A (Statistics
  in Society)}, 163(1):29--38, 2000.

\bibitem{lavori2004dynamic}
Philip~W Lavori and Ree Dawson.
\newblock Dynamic treatment regimes: practical design considerations.
\newblock {\em Clinical trials}, 1(1):9--20, 2004.

\bibitem{petersen2015chapter}
Maya~L Petersen, Joshua Schwab, Elvin~H Geng, and Mark~J van~der Laan.
\newblock Chapter 10: Evaluation of longitudinal dynamic regimes with and
  without marginal structural working models.
\newblock In {\em Adaptive Treatment Strategies in Practice: Planning Trials
  and Analyzing Data for Personalized Medicine}, pages 157--186. Society for
  Industrial and Applied Mathematics, 2015.

\bibitem{benkeser2021improving}
David Benkeser, Iv{\'a}n D{\'\i}az, Alex Luedtke, Jodi Segal, Daniel
  Scharfstein, and Michael Rosenblum.
\newblock Improving precision and power in randomized trials for covid-19
  treatments using covariate adjustment, for binary, ordinal, and time-to-event
  outcomes.
\newblock {\em Biometrics}, 77(4):1467--1481, 2021.

\bibitem{balzertwostage}
Laura~B Balzer, Mark van~der Laan, James Ayieko, Moses Kamya, Gabriel Chamie,
  Joshua Schwab, et~al.
\newblock Two-stage tmle to reduce bias and improve efficiency in cluster
  randomized trials.
\newblock {\em Biostatistics}, 12 2021.

\bibitem{benkeser2020design}
David Benkeser, Keith Horvath, Cathy~J Reback, Joshua Rusow, and Michael
  Hudgens.
\newblock Design and analysis considerations for a sequentially randomized hiv
  prevention trial.
\newblock {\em Statistics in Biosciences}, 12(3):446--467, 2020.

\bibitem{petersen2014causal}
Maya~L Petersen and Mark~J van~der Laan.
\newblock Causal models and learning from data: integrating causal modeling and
  statistical estimation.
\newblock {\em Epidemiology}, 25:418--426, 2014.

\bibitem{van2012targeted}
Mark~J van~der Laan and Susan Gruber.
\newblock Targeted minimum loss based estimation of causal effects of multiple
  time point interventions.
\newblock {\em The International Journal of Biostatistics}, 8(1):1557--4679,
  2012.

\bibitem{bangrobins2005}
H~Bang and J~M Robins.
\newblock Doubly robust estimation in missing data and causal inference models.
\newblock {\em Biometrics}, 61(4):962--973, 2005.

\bibitem{gengestimation}
Elvin~H Geng, Thomas~A Odeny, Rita~E Lyamuya, Alice Nakiwogga-Muwanga, Lameck
  Diero, et~al.
\newblock Estimation of mortality among hiv-infected people on antiretroviral
  treatment in east africa: a sampling based approach in an observational,
  multisite, cohort study.
\newblock {\em Lancet HIV}, 2(3):e107--e116, 2015.

\bibitem{pearl2000}
Judea Pearl.
\newblock {\em Causality: Models, Reasoning, and Inference}.
\newblock Cambridge University Press, 2000.

\bibitem{ltmlepackage}
Samuel~D. Lendle, Joshua Schwab, Maya~L. Petersen, and Mark~J. {van der Laan}.
\newblock {ltmle}: An {R} package implementing targeted minimum loss-based
  estimation for longitudinal data.
\newblock {\em Journal of Statistical Software}, 81(1):1--21, 2017.

\bibitem{petersen2014targeted}
Maya Petersen, Joshua Schwab, Susan Gruber, Nello Blaser, Michael Schomaker,
  and Mark van~der Laan.
\newblock Targeted maximum likelihood estimation for dynamic and static
  longitudinal marginal structural working models.
\newblock {\em Journal of causal inference}, 2(2):147--185, 2014.

\bibitem{van2007super}
Mark~J van~der Laan, Eric~C Polley, and Alan~E Hubbard.
\newblock Super learner.
\newblock {\em Statistical Applications in Genetics and Molecular Biology},
  6(1):1--21, 2007.

\bibitem{cai2020one}
Weixin Cai and Mark~J van~der Laan.
\newblock One-step targeted maximum likelihood estimation for time-to-event
  outcomes.
\newblock {\em Biometrics}, 76(3):722--733, 2020.

\bibitem{van2003unified}
Mark~J van~der Laan and James~M Robins.
\newblock {\em Unified methods for censored longitudinal data and causality}.
\newblock Springer, 2003.

\bibitem{zheng2010asymptotic}
Wenjing Zheng and Mark~J van~der Laan.
\newblock Asymptotic theory for cross-validated targeted maximum likelihood
  estimation.
\newblock {\em U.C. Berkeley Division of Biostatistics Working Paper Series},
  2010.

\bibitem{breiman2017classification}
Leo Breiman, Jerome~H Friedman, Richard~A Olshen, and Charles~J Stone.
\newblock {\em Classification and Regression Trees}.
\newblock Routledge, 1984.

\bibitem{benkeser2016highly}
David Benkeser and Mark van~der Laan.
\newblock The highly adaptive lasso estimator.
\newblock In {\em 2016 IEEE international conference on data science and
  advanced analytics (DSAA)}, pages 689--696. IEEE, 2016.

\bibitem{moore2009covariate}
Kelly~L Moore and Mark~J van~der Laan.
\newblock Covariate adjustment in randomized trials with binary outcomes:
  targeted maximum likelihood estimation.
\newblock {\em Statistics in Medicine}, 28(1):39--64, 2009.

\bibitem{chao2020dynamic}
Yan-Cheng Chao, Howard Trachtman, Debbie~S Gipson, Cathie Spino, Thomas~M
  Braun, and Kelley~M Kidwell.
\newblock Dynamic treatment regimens in small n, sequential, multiple
  assignment, randomized trials: An application in focal segmental
  glomerulosclerosis.
\newblock {\em Contemporary Clinical Trials}, 92:105989, 2020.

\bibitem{tran2019double}
Linh Tran, Constantin Yiannoutsos, Kara Wools-Kaloustian, Abraham Siika, Mark~J
  Van Der~Laan, and Maya Petersen.
\newblock Double robust efficient estimators of longitudinal treatment effects:
  Comparative performance in simulations and a case study.
\newblock {\em The International Journal of Biostatistics}, 15(2):1--27, 2019.

\bibitem{kang2007demystifying}
Joseph~DY Kang and Joseph~L Schafer.
\newblock Demystifying double robustness: A comparison of alternative
  strategies for estimating a population mean from incomplete data.
\newblock {\em Statistical science}, 22(4):523--539, 2007.

\bibitem{robins2007comment}
James Robins, Mariela Sued, Quanhong Lei-Gomez, and Andrea Rotnitzky.
\newblock Comment: Performance of double-robust estimators when" inverse
  probability" weights are highly variable.
\newblock {\em Statistical Science}, 22(4):544--559, 2007.

\bibitem{rotnitzky2012improved}
Andrea Rotnitzky, Quanhong Lei, Mariela Sued, and James~M Robins.
\newblock Improved double-robust estimation in missing data and causal
  inference models.
\newblock {\em Biometrika}, 99(2):439--456, 2012.

\bibitem{gruber2010targeted}
Susan Gruber and Mark~J van~der Laan.
\newblock A targeted maximum likelihood estimator of a causal effect on a
  bounded continuous outcome.
\newblock {\em The International Journal of Biostatistics}, 6(1):1557--4679,
  2010.

\bibitem{ertefaie2020nonparametric}
Ashkan Ertefaie, Nima~S Hejazi, and Mark~J van~der Laan.
\newblock Nonparametric inverse probability weighted estimators based on the
  highly adaptive lasso.
\newblock {\em arXiv preprint arXiv:2005.11303}, 2020.

\end{thebibliography}

\section{Appendix}

\subsection{Appendix A}

In the following, we present algorithmic steps for the IPW estimator, G-computation estimator based on ICEs, and the TMLE. All of these estimators can be implemented with the \emph{ltmle} package \citep{ltmlepackage, petersen2014targeted, tran2019double}. 

Estimates are functions of $P_n$, which is the empirical distribution based on a single sample of size $n$ from $P_0$ that gives each observation weight $\frac{1}{n}$. Here, $\hat{\Psi}_{\tilde{d}}(P_n)$ is an estimate of the true value of an embedded regime $\tilde{d}$, i.e., $\Psi_{\tilde{d}}(P_0)$, and $\hat{\Psi}(P_n)$ is the vector of estimates of the values of a SMART's $D$ embedded regimes. 

\subsubsection{Inverse Probability Weighting (IPW)}
As stated in the main text, the G-computation formula can be re-written as the following IPW estimand:
\[\Psi_{\tilde{d}}(P_0) = \mathbb{E}_0\left[ \frac{\mathbb{I}[\bar{A}(K) = \tilde{\bar{d}}_K(\bar{Z}(K))]}{\prod_{t=1}^K g_0(A(t) \vert \bar{X}(t), \bar{A}(t-1))}Y\right],\]

\noindent where $\prod_{t=1}^K g_0(A(t) = \tilde{d}_{t}(\bar{Z}(t)) \vert \bar{X}(t), \bar{A}(t-1) = \tilde{\bar{d}}_{t-1}(\bar{Z}(t-1)))$ is the product of true time-point-specific predicted probabilities of observed treatment, given the observed treatment and covariate history. 

The IPW estimator is then:

\begin{align*}
\hat{\Psi}_{IPW,\tilde{d}}(P_n) & =\frac{1}{n}\sum_{i=1}^n\frac{\mathbb{I}[\bar{A}_i(K)=\tilde{\bar{d}}_K(\bar{Z}_i(K))]}{\prod_{t=1}^Kg_n(A_i(t)|\bar{X}_i(t), \bar{A}_i(t-1))}Y_i\\
& =\frac{1}{n}\sum_{i=1}^n\mathbb{I}[\bar{A}_i(K)=\tilde{\bar{d}}_K(\bar{Z}_i(K))]\hat{w}_iY_i,
\end{align*}
where $\hat{w}_i = \frac{1}{g_n(A_i(t)|\bar{X}(t), \bar{A}_i(t-1))}$ (i.e., the IPW weights) and $g_n$ is an estimator of the $g_0$ factors. In a SMART, the treatment mechanism is known; thus, if there is no censoring, one could use the known $g_0$ instead of its estimate, or $g_n$ could be a maximum likelihood estimator (MLE) based on a correctly specified lower dimensional parametric model (such as a generalized linear model including either $Z(t)$ alone, or including additional covariates in $X(t)$) -- with or without covariate adjustment. We discuss the efficiency implications of these approaches in Section 4 of the main text.

We note that, while not implemented in the current manuscript, the default for the \emph{ltmle} package is to stabilize these weights via a modified version of the IPW estimator, the modified Horvitz-Thompson estimator: 
\begin{align*}
\hat{\Psi}_{IPW-HT,\tilde{d}}(P_n) & =\frac{\sum_{i=1}^n \frac{\mathbb{I}[\bar{A}_i = \tilde{\bar{d}}_K(\bar{Z}_i(K))]}{\prod_{t=1}^Kg_n(A_i(t)|\bar{X}_i(t), \bar{A}_i(t-1))}Y_i}{\sum_{i = 1}^n \frac{\mathbb{I}[\bar{A}_i = \tilde{\bar{d}}_K(\bar{Z}_i(K))]}{\prod_{t=1}^Kg_n(A_i(t)|\bar{X}_i(t), \bar{A}_i(t-1))}}\\
& =\frac{\sum_{i=1}^n\mathbb{I}[\bar{A}_i(K)=\tilde{\bar{d}}_K(\bar{Z}_i(K))]\hat{w}_iY_i}{\sum_{i=1}^n\mathbb{I}[\bar{A}_i(K) = \tilde{\bar{d}}_K(\bar{Z}_i(K)))]\hat{w}_i} \\
\end{align*}
This is the standard IPW estimator, but divided by the sample average of the weights. Advantages to the Horvitz-Thompson estimator may include a reduction in the variability of the IPW estimates especially in the presence of near-positivity violations, and a containment in the parameter space (for example, if $Y$ is binary, it will ensure an estimate between 0 and 1). The stabilized IPW estimator is equivalent to the unstabilized IPW estimator when the $g_0(A(t)|\bar{Z}(t))$ factors are estimated using the empirical proportions (i.e., a saturated model), in which case the G-computation estimator and TMLE will be equivalent, as well, if the ICE factors are also estimated with saturated regression models (i.e., ``Min. adj. IPW w/ $g_n$" in the simulations).

\subsubsection{G-computation Estimator based on Iterated Conditional Expectations (ICE)}

The above G-computation formula can also be re-written as a series of iterated conditional expectations \citep{bangrobins2005}:
\begin{align*}
    \Psi_d(P_0) = & \mathbb{E}_0[\mathbb{E}_0[ \ldots \\
    & \mathbb{E}_0\left[\mathbb{E}_0\left[Y|\bar{X}(K), \bar{A}(K) = \tilde{\bar{d}}_K(\bar{Z}(K))\right] \vert \bar{X}(K-1), \bar{A}(K-1) = \tilde{\bar{d}}_{K-1}(\bar{Z}(K-1)) \right] \\
    & \ldots \vert X(1), A(1) = \tilde{d}_1(Z(1)) ]]
\end{align*}
To estimate these iterated conditional expectations, one can fit a series of regressions going backwards in time, where each regression uses an updated regression before it (evaluated at the covariate history and treatment rule of interest) as a pseudo-outcome:
\begin{enumerate}
\item At $t = K+1$: Estimate the innermost conditional mean outcome, i.e.: \[Q_{0,K+1}^{\tilde{\bar{d}}_K}= \mathbb{E}_0[Y | \bar{X}(K), \bar{A}(K) = \tilde{\bar{d}}_K(\bar{Z}(K))].\]
\begin{enumerate}
    \item Regress the outcome $Y$ on all past history, i.e., $\bar{X}(K)$ and $\bar{A}(K)$. 
    \item Using the regression in part 1a, predict at the embedded regime of interest $\bar{A}(K) = \tilde{\bar{d}}_K(\bar{Z}(K))$ to obtain \[Q_{n,K+1}^{\tilde{\bar{d}}_K} = \mathbb{E}_n[Y|\bar{X}(K), \bar{A}(K) = \tilde{\bar{d}}_K(\bar{Z}(K))].\]
\end{enumerate}
\item At $t = K$: Estimate second innermost conditional expectation, i.e.:
\begin{align*}
    Q_{0,K}^{\tilde{\bar{d}}_{K-1}}=  \mathbb{E}_0\left[Q_{0,K+1}^{\tilde{\bar{d}}_K}\vert \bar{X}(K-1), \bar{A}(K-1) = \tilde{\bar{d}}_{K-1}(\bar{Z}(K-1)) \right].
\end{align*}
\begin{enumerate}
    \item Regress the pseudo-outcome from the previous step $Q_{n,K+1}^{\tilde{\bar{d}}_K}$ on variables measured before time $K$, i.e., $\bar{X}(K-1)$ and $\bar{A}(K-1)$. 
    \item Using the regression in part 2a, predict at the embedded regime of interest $\bar{A}(K-1)=\tilde{\bar{d}}_{K-1}(\bar{Z}(K-1))$ to obtain 
    \begin{align*}
    Q_{n,K}^{\tilde{\bar{d}}_{K-1}} 
    = \mathbb{E}_n\left[Q_{n,K+1}^{\tilde{\bar{d}}_K}\vert \bar{X}(K-1), \bar{A}(K-1) = \tilde{\bar{d}}_{K-1}(\bar{Z}(K-1)) \right].
\end{align*}
\end{enumerate}
\item Repeat for $t = K-1, \ldots ,3$.
\item At $t = 2$: Estimate the second outermost conditional expectation, i.e.:
\begin{align*}
    Q_{0,2}^{\tilde{d}_1} 
    = & \mathbb{E}_0\left[ \ldots Q_{0,K}^{\tilde{\bar{d}}_{K-1}}  \ldots \vert X(1), A(1) = \tilde{d}_1(Z(1)) \right].
\end{align*}
\begin{enumerate}
    \item Regress the pseudo-outcome from the previous step $Q_{n,3}$ on variables measured before time 2, i.e., $X(1)$ and $A(1)$. 
    \item Using the regression in part 4a, predict at the embedded regime of interest $A(1)=\tilde{d}_1(Z(1))$ to obtain 
    \begin{align*}
   Q_{n,2}^{\tilde{d}_1} 
    = & \mathbb{E}_n\left[ \ldots Q_{n,K}^{\tilde{\bar{d}}_{K-1}}  \ldots \vert X(1), A(1) = \tilde{d}_1(Z(1)) \right].
\end{align*}
\end{enumerate}
\item At $t=1$: Estimate $\Psi_{\tilde{d}}(P_0)$:
\begin{enumerate}
    \item Standardize the pseudo-outcomes from the previous step with respect to the distribution of baseline covariates $X(1)$ by taking the empirical average of $Q_{n,2}^{\tilde{d}_1}$ to obtain $\hat{\Psi}_{\tilde{d},ICE}(P_n)$:
    \begin{align*}
       \hat{\Psi}_{\tilde{d},ICE}(P_n) & = \frac{1}{n}\sum_{i=1}^n  Q_{n,2,i}^{\tilde{d}_1}.
    \end{align*}
\end{enumerate}
\end{enumerate}

\subsubsection{Targeted Maximum Likelihood Estimation (TMLE)}

TMLE combines the treatment mechanisms estimated in the IPW estimator with the ICEs estimated in the G-computation estimator. It is similar to the ICE G-computation estimator procedure, except that each regression at time $t$ is updated before using it as an outcome for the next regression corresponding to time $t-1$. One TMLE procedure is as follows \citep{kang2007demystifying, robins2007comment, rotnitzky2012improved}, noting that outcomes should be transformed (and then back-transformed at the end of the procedure) to be between 0 and 1 \citep{gruber2010targeted}:
\begin{enumerate}
\item At $t = K+1$
\begin{enumerate}
\item Generate an initial estimate of $Q_{0,K+1}^{\tilde{\bar{d}}_K}$ to obtain: \[Q^{0,\tilde{\bar{d}}_K}_{n,K+1}= \mathbb{E}^0_n[Y|\bar{X}(K), \bar{A}(K) = \tilde{\bar{d}}_K(\bar{Z}(K))],\] as in the first step of ICE G-computation estimator.
\item Update $Q^{0,\tilde{\bar{d}}_K}_{n,K+1}$ to $Q^{*,\tilde{\bar{d}}_K}_{n,K+1}$ as follows:
\begin{enumerate}
    \item Fit a logistic regression of $Y$ on the intercept using the logit of $Q^{0,\tilde{\bar{d}}_K}_{n,K+1}$ as offset and weight \\
    $\mathbb{I}[\bar{A}(K)=\tilde{\bar{d}}_K(\bar{Z}(K))]/\prod_{t=1}^K g_n(A(t) \vert \bar{X}(t),\bar{A}(t-1))$.
    \item Obtain predicted probabilities using the logistic regression fit in step 1b at $\bar{A}(K) = \tilde{\bar{d}}_K(\bar{Z}(K))$ to obtain the targeted estimate $Q^{*,\tilde{\bar{d}}_K}_{n,K+1}$.
\end{enumerate}
\end{enumerate}
\item At $t = K$
\begin{enumerate}
\item Using $Q^{*,\tilde{\bar{d}}_K}_{n,K+1}$ from the previous step as a pseudo-outcome, generate an initial estimate of $Q^{\tilde{\bar{d}}_{K-1}}_{0,K}$: 
\begin{align*}
    Q^{0,\tilde{\bar{d}}_{K-1}}_{n,K}
    = \mathbb{E}^0_n[Q^{*,\tilde{\bar{d}}_K}_{n,K+1}|\bar{X}(K-1), \bar{A}(K-1) = \tilde{\bar{d}}_{K-1}(\bar{Z}(K-1))].
\end{align*}
\item Update $Q^{0,\tilde{\bar{d}}_{K-1}}_{n,K}$ to $Q^{*,\tilde{\bar{d}}_{K-1}}_{n,K}$ as follows:
\begin{enumerate}
    \item Fit a logistic regression of $Q^{*,\tilde{\bar{d}}_K}_{n,K+1}$ on the intercept using the logit of $Q^{0,\tilde{\bar{d}}_{K-1}}_{n,K}$ as offset and weight\\ $\mathbb{I}[\bar{A}(K-1) =\tilde{\bar{d}}_{K-1}(\bar{Z}(K-1))]/\prod_{t=1}^{K-1} g_n(A(t) \vert \bar{X}(t), \bar{A}(t-1)).$
    \item Obtain predicted probabilities using the logistic regression fit in step 2b at $\bar{A}(K-1) = \tilde{\bar{d}}_{K-1}(\bar{Z}(K-1)))$ to obtain the targeted estimate $Q^{*,\tilde{\bar{d}}_{K-1}}_{n,K}$.
\end{enumerate}
\end{enumerate}
\item Repeat for $t = K-1, \ldots ,3$.
\item At $t = 2$
\begin{enumerate}
\item Using $Q^{*,\tilde{\bar{d}}_2}_{n,3}$ from the previous step as a pseudo-outcome, generate an initial estimate of $Q^{\tilde{d}_{1}}_{0,2}$: 
\begin{align*}
    Q^{0,\tilde{d}_{1}}_{n,2} = \mathbb{E}^0_n[Q^{*,\tilde{\bar{d}}_2}_{n,3}|X(1), A(1) = \tilde{d}_{1}(Z(1))].
\end{align*}
\item Update $Q^{0,\tilde{d}_{1}}_{n,2}$ to $Q^{*,\tilde{d}_{1}}_{n,2}$ as follows:
\begin{enumerate}
    \item Fit a logistic regression of $Q^{*,\tilde{\bar{d}}_2}_{n,3}$ on the intercept using the logit of $Q^{0,\tilde{d}_{1}}_{n,2}$ as offset and weight $\mathbb{I}[A(1) =\tilde{d}_{1}(Z(1))]/g_n(A(1) \vert X(1)).$
    \item Obtain predicted probabilities using the logistic regression fit in step 4b at $A(1) = \tilde{d}_{1}(Z(1))$ to obtain the targeted estimate $Q^{*,\tilde{d}_1}_{n,2}$.
\end{enumerate}
\end{enumerate}
\item At $t=1$: Estimate $\Psi_{\tilde{d}}(P_0)$.
\begin{enumerate}
    \item Standardize the pseudo-outcomes from the previous step with respect to the distribution of baseline covariates $X(1)$ by taking the empirical average of $Q^{*,\tilde{d}_1}_{n,2}$ to obtain $\hat{\Psi}_{\tilde{d},TMLE}(P_n)$:
        \begin{align*}
       \hat{\Psi}_{\tilde{d},TMLE}(P_n) & = \frac{1}{n}\sum_{i=1}^n  Q^{*,\tilde{d}_1}_{n,2,i}.
    \end{align*}
\end{enumerate}
\end{enumerate}
Similar to the IPW estimator, in a SMART, the treatment mechanism $g_0$ is known, so one could estimate the $g$ factors using an MLE based on a correctly specified model, or one could use the true $g_0$ values. The ICE factors in the TMLE procedure are recommended to be estimated using flexible machine learning algorithms, such as SuperLearner \citep{van2007super}.

\subsection{Appendix B}\label{inferenceone}

Below, we discuss technical details for: 1) individual inference around an estimate of one embedded regime value; and 2) simultaneous inference for all embedded regimes values.

\subsubsection{Inference for the Value of One Embedded Regime}

Using influence-curve based inference, we describe how to construct 95\% confidence intervals around a TMLE estimate with nominal to conservative coverage for the value of one embedded regime.

An estimator $\hat{\Psi}_{\tilde{d}}$ is asymptotically linear for its true value $\Psi_{\tilde{d}}(P_0)$ if $\hat{\Psi}_{\tilde{d}}(P_n) - \Psi_{\tilde{d}}(P_0) = \frac{1}{n}\sum_{i=1}^n IC_{\tilde{d}}(P_0) + o_P(n^{-1/2})$,
where $IC_{\tilde{d}}$ is the estimator's influence curve. This translates to the limit distribution $n^{1/2}(\hat{\Psi}_{\tilde{d}}(P_n) - \Psi_{\tilde{d}}(P_0)) \overset{d}{\to} N(0, \sigma^2_{\tilde{d},0})$, allowing an estimate of $\sigma^2_{\tilde{d},0}$ to be used to construct a Wald-type confidence interval. 

TMLE's influence curve is the efficient influence curve of the target statistical parameter; i.e., at a distribution $P$, its influence curve is $IC_{\tilde{d}}^*(P)(O) = \sum_{t=1}^{K}IC_{\tilde{d}_t}^*(Q, g)(O) +  Q^{\tilde{d}_1}_{2} - \Psi_{\tilde{d}}(P)$, where: $IC^*_{\tilde{d}_{K}}(O) = \frac{\mathbb{I}[\bar{A}(K) = \tilde{\bar{d}}_K(\bar{Z}(K))](Y - Q^{\tilde{\bar{d}}_K}_{K+1})}{\prod_{t=1}^K g(A(t) \vert \bar{X}(t), \bar{A}(t-1))}$,
and, for $t = 1,...,K-1$, $IC^*_{\tilde{d}_t}(O) = \frac{\mathbb{I}[\bar{A}(t) = \tilde{\bar{d}}_t(\bar{Z}(t))](Q^{\tilde{\bar{d}}_{t+1}}_{t+2} - Q^{\tilde{\bar{d}}_t}_{t+1})}{\prod_{j=1}^t g(A(j) \vert \bar{X}(j), \bar{A}(j-1))}$ \citep{van2012targeted, bangrobins2005}.

We present this because the sample variance of $\hat{IC}^*_{\tilde{d}}(P_n)$ can be used as an estimate of $\sigma^2_{\tilde{d},0}$, i.e., $\hat{\sigma}^2_{\tilde{d}} = \frac{1}{n}\sum_i^n \hat{IC}^*_{\tilde{d}}(O_i)^2$. Thus, for a TMLE estimator $\hat{\Psi}_{\tilde{d}}$ and its corresponding working influence curve estimate $\hat{IC}^*_{\tilde{d}}$, we obtain conservative (or nominal) inference on the value of one embedded regime $d$ by constructing a 95\% confidence interval like so: $\hat{\Psi}_{\tilde{d}}(P_n) \pm \Phi^{-1}(0.975)\frac{\hat{\sigma}_{\tilde{d}}}{\sqrt{n}}$.

It is straightforward to extend this approach to inference for contrasts of pairs of embedded regimes $\tilde{d}^{(i)}$ and $\tilde{d}^{(j)}$ using the functional delta method. For example, the efficient influence curve corresponding to the contrast of the values $\Psi_{\tilde{d}^{(i)}}(P_0) - \Psi_{\tilde{d}^{(j)}}(P_0)$ is $IC_{\tilde{d}^{(i)}}^{*}(O) - IC_{\tilde{d}^{(j)}}^{*}(O)$.

\subsubsection{Simultaneous Inference for all Embedded Regimes}\label{inferencesimult}

Let $\hat{\Psi}(P_n)$ be an estimator of $\Psi(P_0)$, the vector of true values of the $D$ embedded regimes within a SMART. This estimator is asymptotically linear with influence curve matrix $IC(P_0)\in \mathbb{R}^D$ if $\hat{\Psi}(P_n) - \Psi(P_0) = \frac{1}{n}\sum_{i=1}^n IC(P_0) + o_P(n^{-1/2})$,
which implies $n^{1/2}\{\hat{\Psi}(P_n) - \Psi(P_0)\}  \overset{d}{\to} \mathcal{N}(0, \Sigma_0)$. Here, $\Sigma_0$ is the $D \times D$ covariance matrix of the influence curve matrix, i.e., $\Sigma_0 = \mathbb{E}_0[IC(P_0)IC(P_0)^T(O)]$, which can be estimated empirically; thus, the covariance matrix of $\hat{\Psi}(P_n)$ can be estimated with $\Sigma_n/n$. Let $\rho = Corr[IC(P_0)]$. Then a simultaneous confidence interval for one of the SMART's embedded regimes $\tilde{d}^{(j)}$, $j = 1,...,D$ is: $\hat{\Psi}_{\tilde{d}^{(j)}}(P_n) \pm q_{0.95} \sqrt{\frac{\Sigma_n(j,j)}{n}}$, where $q_{0.95}$ is the $95^{th}$ quantile of $\max_j \vert Z(j) \vert$ for $Z \sim \mathcal{N}(0, \rho)$. In this way, under the above conditions for asymptotic linearity in the previous section, the probability that all $D$ confidence intervals capture their respective truths is 95\%, i.e.: $P\left(\Psi_{\tilde{d}^{(j)}}(P_0) \in \hat{\Psi}_{\tilde{d}^{(j)}}(P_n) \pm q_{0.95} \sqrt{\frac{\Sigma_n(j,j)}{n}}\right) = 0.95$. 
We note that $\Sigma_n(j,j)$ is equivalent to one $\hat{\sigma}^2_{\tilde{d}}$ from the previous section. We refer the reader to \cite{cai2020one} for more background of simultaneous confidence intervals for multivariate normally distributed random variables within the TMLE framework.

\subsection{Appendix C}

\subsubsection{SuperLearner and ltmle Configurations}

When SuperLearner was employed in G-computation and TMLE to estimate the ICE factors, we used 10-fold cross-validation and the following library (the default library for \emph{ltmle}): generalized linear models (SL.glm; with and without correlation screening algorithm), stepwise forward and backward regression by AIC (SL.stepAIC; with and without correlation screening algorithm),  Bayes' generalized linear model (SL.bayesglm; with and without correlation screening algorithm), stepwise forward regression by AIC (SL.step.forward; with correlation screening algorithm), and stepwise forward and backward regression with pairwise interactions (SL.step.interaction; with correlation screening algorithm). 

Appendix D shows additional simulations that examine the performance results under additional flexible SuperLearner libraries including tree-based algorithms \citep{breiman2017classification} and the Highly Adaptive Lasso (HAL) \citep{benkeser2016highly, ertefaie2020nonparametric}.

We used the default ltmle function arguments (e.g., \textit{gbounds}, \textit{Yrange}, \textit{stratify}), except for the following. First, the ones the user must specify -- i.e.,  \textit{data} (including the baseline covariates, Stage 1 treatment, intermediate covariates, and outcome), \textit{Anodes} ($A(1)$ and $A(2)$), \textit{Lnodes} (all baseline and time-varying covariates, i.e., all variables in \textit{data} that were not \textit{Anodes} and \textit{Ynodes}),  \textit{Ynodes} ($Y$), and \textit{abar} (a matrix, of dimension $n$ by number of \textit{Anodes}, of the desired embedded regime intervention). Second, we used the following additional arguments: \textit{deterministic.g.function} (needed to encode the ADAPT data's treatment mechanism process so that those who were initially in SOC and did not have a lapse were deterministically assigned to SOC outreach for their Stage 2 treatment), \textit{deterministic.Q.function} (needed to encode that if a person died, transferred, or withdrew, then they would deterministically receive their observed outcome), \textit{gform} (to ensure that the treatment mechanism was only a function of the necessary covariates that fulfill the sequential randomization condition, as opposed to all covariates, which could result in overfitting), and \textit{SL.library} (which was set to ltmle's default SuperLearner library [described above]).

\subsubsection{Fully Simulated, Simple SMART Design (DGP 1)}

Mimicking a simple SMART with 8 embedded regimes in which re-randomization is based only on intermediate covariates, the covariates, treatments and outcome were generated as follows:

\begin{align*}
    X(1) &\sim Normal(\mu = 0,\sigma = 1) \\
    A(1) & = Bernoulli(p = 0.5) \text{ with support \{0,1\}}\\
    L(2) &\sim Bernoulli(p = \textrm{expit}(X(1) + A(1)) \\
    S(2) &\sim Normal(\mu = X(1) + 2A(1), \sigma = 1) \\
    A(2) & \sim \begin{cases}
        \text{If } L(2) = 1,   Bern(p = 0.5) \text{ with support \{1,2\}}\\ 
        \text{If } L(2) = 0,   Bern(p = 0.5) \text{ with support \{3,4\}}\\
        \end{cases} \\
    Y &\sim Bernoulli(p = Q(X(1), A(1), X(2), A(2)) \\
     & =  Bernoulli(p = \logit^{-1}(\logit(c) + S(2) + 0.5X(1)^2 + \log(|X1|+.01))))
\end{align*}
where $c = 1 - (.28, .26, .28, .3, .29, .3, .21, .2)$, a vector of fixed constants unique to each of the 8 embedded regimes (respectively). The true causal values of each embedded regime are given in Table \ref{tableDGP1truevals}.

For this DGP, the minimal adjustment set (i.e., the minimum set of covariates needed to ensure that the statistical estimand was equal to the causal effect of interest) included only the time-varying covariate $\bar{Z}(2) = L(2)$; we also considered a full adjustment set that included the baseline covariate $X(1)$ and time varying covariates $X(2) = (L(2), S(2))$. The marginal probability of the occurrence of the outcome was 74.4\%. 

\begin{table}[]
\begin{tabular}{|l|l|l|l|l|}
\hline
Causal Parameter & $A(1)$ & $A(2)$ if $L(2)=1$ & $A(2)$ if $L(2)=0$ & True Value \\ \hline
$\Psi_{\tilde{d}^1}(P_{U,X})$ & 0 & 1 & 3 & 0.6061 \\ \hline
$\Psi_{\tilde{d}^2}(P_{U,X})$ & 1 & 1 & 3 & 0.8634 \\ \hline
$\Psi_{\tilde{d}^3}(P_{U,X})$ & 0 & 2 & 3 & 0.6060 \\ \hline
$\Psi_{\tilde{d}^4}(P_{U,X})$ & 1 & 2 & 3 & 0.8517 \\ \hline
$\Psi_{\tilde{d}^5}(P_{U,X})$ & 0 & 1 & 4 & 0.6420 \\ \hline
$\Psi_{\tilde{d}^6}(P_{U,X})$ & 1 & 1 & 4 & 0.8777 \\ \hline
$\Psi_{\tilde{d}^7}(P_{U,X})$ & 0 & 2 & 4 & 0.6421 \\ \hline
$\Psi_{\tilde{d}^8}(P_{U,X})$ & 1 & 2 & 4 & 0.8660 \\ \hline

\end{tabular}
\label{tableDGP1truevals}
\caption{True causal parameter values of each of the embedded regimes in DGP 1.}
\end{table}

\subsubsection{Outcome-blind Simulation Based on ADAPT-R (DGP 2)}

DGP 2 mimicked the way ADAPT-R data were generated. The covariates $X(1)$ (baseline covariates) and $X(2)$ (which included $L(2)$, whether there was a lapse in care, $D(2)$, whether a person died before $t=2$, $M(2)$, whether a person transferred or withdrew before $t=2$, and the other time varying covariates listed in Section 3.3.1) were sampled, with replacement, from the empirical distributions of the actual ADAPT-R data. Next, we randomly generated the treatments and outcome as follows:
        \begin{align*}
           A(1) \sim & Multinom(1, p_{SMS} = p_{CCT} = p_{SOC} = 1/3) \\
           & \text{ with support \{SMS, CCT, SOC\}}\\
         A(2) & \begin{cases}
        \sim Multinom(1, p_{SMS+CCT} = p_{Nav.} = p_{SOC} = 1/3)\\
        \indent \text{with support \{SMS+CCT, Nav., SOC\}}, \text{ if } L(2) = 1\\ 
        \sim Bernoulli(p_{discont.} = p_{cont.} = 1/2)\\
        \indent \text{with support \{Discont., Cont.\} }, \text{ if } L(2) = 0 \text{ and } A(1) \in \{\text{SMS, CCT}\}\\ 
        = \text{Continue if } L(2) = 0 \text{ and } A(1) = \text{SOC}\\ 
        = \text{Missing if } $D(2) = 1$ \text{ or } $M(2) = 1$ 
        \end{cases}\\
        Y & \begin{cases}
        \sim Bernoulli(p = Q_1(X(1), A(1))) \text{ if } M(1)=1 \text{ and } D(2) = 0\\ 
        \sim Bernoulli(p = Q_2(X(1), A(1), X(2), A(2))) \text{ if } M(1)=0 \text{ and } D(2) = 0\\  
        = 0 \text{ if } D(2) = 1 \\ 
        \end{cases}
        \end{align*}
Here, $Q_1(X(1), A(1))) = .35 + (X(1)_{\text{alcohol}} = \text{level 1 and 2})*.1$, $Q_2(X(1), A(1), X(2), A(2))) = \logit^{-1}(\logit(c) + L(2) + X(2)_{\text{time to re-randomization}}/300 - \mathbb{I}[X(1)_{\text{sex}} = male)*(X(1)_{\text{age}}/10))$, and c = 1 - (.28, .26, .28, .3, .29, .3, .21, .2, .21, .18, .18, .22, .13, .22) (a vector of fixed constants respective to each of the 15 regimes). The true causal values of each embedded regime are listed in Table \ref{tableDGP2truevalues}.

We highlight that if a person died before $t=2$ (i.e., $D(2)=1$), that participant had missing time-updated covariates and second-stage treatment, and deterministically received a value of 0 (viral suppression failure) for the outcome. Additionally, if a person transferred or withdrew before $t=2$ (i.e., $M(2)=1$), the participant also had missing time-updated covariates and second-stage treatment; however, the outcome (remaining alive and unsuppressed at year 2) was still measured on these participants. These restrictions on the dependence between past variables and the outcome were encoded using the \emph{deterministic.Q.function} argument for the \emph{ltmle} function; namely, that if a person died, transferred, or withdrew (denoted by $D(2)=1$ or $M(2)=1$), then the person deterministically received the actual $Y$ value present the data. Similarly, the fact that participants who were initially randomized to SOC and did not have a lapse in care were deterministically assigned SOC as their second-stage treatment was encoded via \emph{ltmle}'s  \emph{deterministic.g.function} argument. This ensures that those who initially received SOC and succeeded in the first year were not re-randomized to either continue \emph{or} discontinue SOC (as with those who received SMS or CCT initially) -- instead, they were deterministically given SOC outreach.

The minimal covariate adjustment set for this DGP included $\bar{Z}(2) = (A(1), L(2))$, $M(2)$, and $D(2)$. We also considered a full adjustment set that included all baseline and time-varying covariates in $X(1)$ and $X(2)$. The marginal probability of the occurrence of the outcome was simulated to be 72.0\%, chosen to be similar to ADAPT-R's actual probability of viral suppression at 2 years.

\begin{table}[]
\begin{tabular}{|l|l|l|l|l|}
\hline
Causal Parameter & $A(1)$ & $A(2)$ if $L(2)=1$ & $A(2)$ if $L(2)=0$ & True Value \\ \hline
$\Psi_{\tilde{d}_1}(P_{U,X})$ & SOC & SOC & Continue & 0.7376 \\ \hline
$\Psi_{\tilde{d}_2}(P_{U,X})$ & SMS & SOC & Continue & 0.6884 \\ \hline
$\Psi_{\tilde{d}_3}(P_{U,X})$ & CCT & SOC & Continue & 0.6884 \\ \hline
$\Psi_{\tilde{d}_4}(P_{U,X})$ & SOC & SMS + CCT & Continue & 0.7314 \\ \hline
$\Psi_{\tilde{d}_5}(P_{U,X})$ & SMS & SMS + CCT & Continue & 0.6814 \\ \hline
$\Psi_{\tilde{d}_6}(P_{U,X})$ & CCT & SMS + CCT & Continue & 0.6814 \\ \hline
$\Psi_{\tilde{d}_7}(P_{U,X})$ & SOC & Navigator & Continue & 0.7285 \\ \hline
$\Psi_{\tilde{d}_8}(P_{U,X})$ & SMS & Navigator & Continue & 0.6795 \\ \hline
$\Psi_{\tilde{d}_9}(P_{U,X})$ & CCT & Navigator & Continue & 0.6795 \\ \hline
$\Psi_{\tilde{d}_{10}}(P_{U,X})$ & SMS & SOC & Discontinue & 0.7080 \\ \hline
$\Psi_{\tilde{d}_{11}}(P_{U,X})$ & CCT & SOC & Discontinue & 0.7080 \\ \hline
$\Psi_{\tilde{d}_{12}}(P_{U,X})$ & SMS & SMS + CCT & Discontinue & 0.7010 \\ \hline
$\Psi_{\tilde{d}_{13}}(P_{U,X})$ & CCT & SMS + CCT & Discontinue & 0.7010 \\ \hline
$\Psi_{\tilde{d}_{14}}(P_{U,X})$ & SMS & Navigator & Discontinue & 0.6991 \\ \hline
$\Psi_{\tilde{d}_{15}}(P_{U,X})$ & CCT & Navigator & Discontinue & 0.6991 \\ \hline
\end{tabular}
\caption{True causal parameter values of each of the embedded regimes in DGP 2.}
\end{table}
\label{tableDGP2truevalues}

\begin{table}[]
\begin{tabular}{|l|l|l|llll|}
\hline
Rule & Estimator & Bias & \multicolumn{1}{l|}{Var.} & \multicolumn{1}{l|}{C.I. Width} & \multicolumn{1}{l|}{Ind. Cov. (\%)} & Simult. Cov. (\%) \\ \hline
\multirow{5}{*}{1} & Min. adj. IPW (w/ $g_0$) & 0.0001 & \multicolumn{1}{l|}{0.0013} & \multicolumn{1}{l|}{0.1366} & \multicolumn{1}{l|}{94.8} & 94.9 \\ \cline{2-7} 
 & Min. adj. IPW (w/ $g_n$) & 0.0006 & \multicolumn{1}{l|}{0.0006} & \multicolumn{1}{l|}{0.1367} & \multicolumn{1}{l|}{99.6} & 99.9 \\ \cline{2-7} 
 & Full adj. IPW & 0.0003 & \multicolumn{1}{l|}{0.0006} & \multicolumn{1}{l|}{0.1371} & \multicolumn{1}{l|}{99.6} & 99.9 \\ \cline{2-7} 
 & Full adj. G-comp. & 0.0009 & \multicolumn{4}{c|}{N/A} \\ \cline{2-7} 
 & Full adj. TMLE & 0.0003 & \multicolumn{1}{l|}{0.0005} & \multicolumn{1}{l|}{0.0871} & \multicolumn{1}{l|}{93.6} & 94.4 \\ \hline
\multirow{5}{*}{2} & Min. adj. IPW (w/ $g_0$) & 0.0011 & \multicolumn{1}{l|}{0.0013} & \multicolumn{1}{l|}{0.1398} & \multicolumn{1}{l|}{94.3} & 94.9 \\ \cline{2-7} 
 & Min. adj. IPW (w/ $g_n$) & 0.0006 & \multicolumn{1}{l|}{0.0005} & \multicolumn{1}{l|}{0.1402} & \multicolumn{1}{l|}{99.8} & 99.9 \\ \cline{2-7} 
 & Full adj. IPW & 0.0007 & \multicolumn{1}{l|}{0.0005} & \multicolumn{1}{l|}{0.1404} & \multicolumn{1}{l|}{99.7} & 99.9 \\ \cline{2-7} 
 & Full adj. G-comp. & 0.0038 & \multicolumn{4}{c|}{N/A} \\ \cline{2-7} 
 & Full adj. TMLE & 0.0007 & \multicolumn{1}{l|}{0.0005} & \multicolumn{1}{l|}{0.0860} & \multicolumn{1}{l|}{94.6} & 94.4 \\ \hline
\multirow{5}{*}{3} & Min. adj. IPW (w/ $g_0$) & 0.0002 & \multicolumn{1}{l|}{0.0012} & \multicolumn{1}{l|}{0.1366} & \multicolumn{1}{l|}{94.5} & 94.9 \\ \cline{2-7} 
 & Min. adj. IPW (w/ $g_n$) & 0.0007 & \multicolumn{1}{l|}{0.0006} & \multicolumn{1}{l|}{0.1367} & \multicolumn{1}{l|}{99.3} & 99.9 \\ \cline{2-7} 
 & Full adj. IPW & 0.0008 & \multicolumn{1}{l|}{0.0005} & \multicolumn{1}{l|}{0.1370} & \multicolumn{1}{l|}{99.6} & 99.9 \\ \cline{2-7} 
 & Full adj. G-comp. & 0.0022 & \multicolumn{4}{c|}{N/A} \\ \cline{2-7} 
 & Full adj. TMLE & 0.0007 & \multicolumn{1}{l|}{0.0005} & \multicolumn{1}{l|}{0.0871} & \multicolumn{1}{l|}{94.9} & 94.4 \\ \hline
\multirow{5}{*}{4} & Min. adj. IPW (w/ $g_0$) & 0.0015 & \multicolumn{1}{l|}{0.0013} & \multicolumn{1}{l|}{0.1398} & \multicolumn{1}{l|}{93.8} & 94.9 \\ \cline{2-7} 
 & Min. adj. IPW (w/ $g_n$) & 0.0008 & \multicolumn{1}{l|}{0.0005} & \multicolumn{1}{l|}{0.1402} & \multicolumn{1}{l|}{99.8} & 99.9 \\ \cline{2-7} 
 & Full adj. IPW & 0.0012 & \multicolumn{1}{l|}{0.0005} & \multicolumn{1}{l|}{0.1404} & \multicolumn{1}{l|}{100.0} & 99.9 \\ \cline{2-7} 
 & Full adj. G-comp. & 0.0025 & \multicolumn{4}{c|}{N/A} \\ \cline{2-7} 
 & Full adj. TMLE & 0.0012 & \multicolumn{1}{l|}{0.0005} & \multicolumn{1}{l|}{0.0860} & \multicolumn{1}{l|}{95.3} & 94.4 \\ \hline
\multirow{5}{*}{5} & Min. adj. IPW (w/ $g_0$) & 0.0010 & \multicolumn{1}{l|}{0.0016} & \multicolumn{1}{l|}{0.1567} & \multicolumn{1}{l|}{95.1} & 94.9 \\ \cline{2-7} 
 & Min. adj. IPW (w/ $g_n$) & 0.0004 & \multicolumn{1}{l|}{0.0003} & \multicolumn{1}{l|}{0.1572} & \multicolumn{1}{l|}{100.0} & 99.9 \\ \cline{2-7} 
 & Full adj. IPW & 0.0001 & \multicolumn{1}{l|}{0.0003} & \multicolumn{1}{l|}{0.1575} & \multicolumn{1}{l|}{100.0} & 99.9 \\ \cline{2-7} 
 & Full adj. G-comp. & 0.0042 & \multicolumn{4}{c|}{N/A} \\ \cline{2-7} 
 & Full adj. TMLE & 0.0002 & \multicolumn{1}{l|}{0.0003} & \multicolumn{1}{l|}{0.0632} & \multicolumn{1}{l|}{94.1} & 94.4 \\ \hline
\multirow{5}{*}{6} & Min. adj. IPW (w/ $g_0$) & 0.0001 & \multicolumn{1}{l|}{0.0016} & \multicolumn{1}{l|}{0.1577} & \multicolumn{1}{l|}{94.2} & 94.9 \\ \cline{2-7} 
 & Min. adj. IPW (w/ $g_n$) & 0.0001 & \multicolumn{1}{l|}{0.0003} & \multicolumn{1}{l|}{0.1581} & \multicolumn{1}{l|}{100.0} & 99.9 \\ \cline{2-7} 
 & Full adj. IPW & 0.0001 & \multicolumn{1}{l|}{0.0003} & \multicolumn{1}{l|}{0.1584} & \multicolumn{1}{l|}{100.0} & 99.9 \\ \cline{2-7} 
 & Full adj. G-comp. & 0.0065 & \multicolumn{4}{c|}{N/A} \\ \cline{2-7} 
 & Full adj. TMLE & 0.0002 & \multicolumn{1}{l|}{0.0003} & \multicolumn{1}{l|}{0.0604} & \multicolumn{1}{l|}{93.9} & 94.4 \\ \hline
\multirow{5}{*}{7} & Min. adj. IPW (w/ $g_0$) & 0.0004 & \multicolumn{1}{l|}{0.0017} & \multicolumn{1}{l|}{0.1561} & \multicolumn{1}{l|}{94.9} & 94.9 \\ \cline{2-7} 
 & Min. adj. IPW (w/ $g_n$) & 0.0005 & \multicolumn{1}{l|}{0.0003} & \multicolumn{1}{l|}{0.1564} & \multicolumn{1}{l|}{100.0} & 99.9 \\ \cline{2-7} 
 & Full adj. IPW & 0.0004 & \multicolumn{1}{l|}{0.0003} & \multicolumn{1}{l|}{0.1566} & \multicolumn{1}{l|}{100.0} & 99.9 \\ \cline{2-7} 
 & Full adj. G-comp. & 0.0069 & \multicolumn{4}{c|}{N/A} \\ \cline{2-7} 
 & Full adj. TMLE & 0.0003 & \multicolumn{1}{l|}{0.0003} & \multicolumn{1}{l|}{0.0653} & \multicolumn{1}{l|}{94.9} & 94.4 \\ \hline
\multirow{5}{*}{8} & Min. adj. IPW (w/ $g_0$) & 0.0014 & \multicolumn{1}{l|}{0.0016} & \multicolumn{1}{l|}{0.1571} & \multicolumn{1}{l|}{94.7} & 94.9 \\ \cline{2-7} 
 & Min. adj. IPW (w/ $g_n$) & 0.0010 & \multicolumn{1}{l|}{0.0003} & \multicolumn{1}{l|}{0.1573} & \multicolumn{1}{l|}{100.0} & 99.9 \\ \cline{2-7} 
 & Full adj. IPW & 0.0007 & \multicolumn{1}{l|}{0.0003} & \multicolumn{1}{l|}{0.1576} & \multicolumn{1}{l|}{100.0} & 99.9 \\ \cline{2-7} 
 & Full adj. G-comp. & 0.0046 & \multicolumn{4}{c|}{N/A} \\ \cline{2-7} 
 & Full adj. TMLE & 0.0008 & \multicolumn{1}{l|}{0.0003} & \multicolumn{1}{l|}{0.0626} & \multicolumn{1}{l|}{94.5} & 94.4 \\ \hline
\end{tabular}
\caption{(Caption on the following page.)}
\label{DGP1table}
\end{table}

\begin{table}
\contcaption{DGP 1: Performance (absolute bias [Abs. Bias], variance [Var.], confidence interval [C.I.] width, individual [Ind.] and simultaneous [Simult.] confidence interval coverage [Cov.]) of each of the 5 estimators for the values of a simple sequential multiple assignment randomized trial's (SMART's) 8 embedded regimes. The 5 estimators evaluated are: 1) an inverse probability weighted (IPW) estimator with weights based on the true, known probability of receiving treatment (``Min. adj IPW (w/ $g_0$)"); 2) an IPW estimator with estimated weights based on the empirical proportion of receiving treatment, which is equivalent to a TMLE or G-computation estimator where iterated conditional expectation (ICE) factors are estimated with saturated regression models non-parametric maximum likelihood estimator (``Min. adj IPW (w/ $g_n$)"); 3) an IPW estimator with estimated weights that adjust for all covariates (``Full adj. IPW"); 4) a G-computation estimator based on ICEs estimated with machine learning that adjust for all covariates (``Full adj. G-comp."); and 5) a targeted maximum likelihood estimator (TMLE) that adjusts for all covariates (``Full adj. TMLE").}
\end{table}

\begin{sidewaystable}[]
\scalebox{.65}{
\begin{tabular}{llllllll|l|l|l|llll|}
\cline{1-7} \cline{9-15}
\multicolumn{1}{|l|}{Rule} & \multicolumn{1}{l|}{Estimator} & \multicolumn{1}{l|}{Abs. Bias} & \multicolumn{1}{l|}{Var.} & \multicolumn{1}{l|}{C.I. Width} & \multicolumn{1}{l|}{Ind. Cov. (\%)} & \multicolumn{1}{l|}{Simult. Cov. (\%)} &  & Rule & Estimator & Abs. Bias & \multicolumn{1}{l|}{Var.} & \multicolumn{1}{l|}{C.I. Width} & \multicolumn{1}{l|}{Ind. Cov. (\%)} & Simult. Cov. (\%) \\ \cline{1-7} \cline{9-15} 
\multicolumn{1}{|l|}{\multirow{5}{*}{1}} & \multicolumn{1}{l|}{Min. adj. IPW (w/ $g_0$)} & \multicolumn{1}{l|}{0.0028} & \multicolumn{1}{l|}{0.0023} & \multicolumn{1}{l|}{0.1880} & \multicolumn{1}{l|}{93.6} & \multicolumn{1}{l|}{92.7} &  & \multirow{5}{*}{8} & Min. adj. IPW (w/ $g_0$) & 0.0005 & \multicolumn{1}{l|}{0.0014} & \multicolumn{1}{l|}{0.1488} & \multicolumn{1}{l|}{94.8} & 92.7 \\ \cline{2-7} \cline{10-15} 
\multicolumn{1}{|l|}{} & \multicolumn{1}{l|}{Min. adj. IPW (w/ $g_n$)} & \multicolumn{1}{l|}{0.0015} & \multicolumn{1}{l|}{0.0008} & \multicolumn{1}{l|}{0.1892} & \multicolumn{1}{l|}{99.8} & \multicolumn{1}{l|}{100.0} &  &  & Min. adj. IPW (w/ $g_n$) & 0.0001 & \multicolumn{1}{l|}{0.0005} & \multicolumn{1}{l|}{0.1494} & \multicolumn{1}{l|}{99.9} & 100.0 \\ \cline{2-7} \cline{10-15} 
\multicolumn{1}{|l|}{} & \multicolumn{1}{l|}{Full adj. IPW} & \multicolumn{1}{l|}{0.0028} & \multicolumn{1}{l|}{0.0007} & \multicolumn{1}{l|}{0.1962} & \multicolumn{1}{l|}{100.0} & \multicolumn{1}{l|}{100.0} &  &  & Full adj. IPW & 0.0008 & \multicolumn{1}{l|}{0.0005} & \multicolumn{1}{l|}{0.1567} & \multicolumn{1}{l|}{99.9} & 100.0 \\ \cline{2-7} \cline{10-15} 
\multicolumn{1}{|l|}{} & \multicolumn{1}{l|}{Full adj. G-comp.} & \multicolumn{1}{l|}{0.0147} & \multicolumn{4}{c|}{N/A} &  &  & Full adj. G-comp. & 0.0112 & \multicolumn{4}{c|}{N/A} \\ \cline{2-7} \cline{10-15} 
\multicolumn{1}{|l|}{} & \multicolumn{1}{l|}{Full adj. TMLE} & \multicolumn{1}{l|}{0.0018} & \multicolumn{1}{l|}{0.0005} & \multicolumn{1}{l|}{0.0872} & \multicolumn{1}{l|}{94.9} & \multicolumn{1}{l|}{92.6} &  &  & Full adj. TMLE & 0.0005 & \multicolumn{1}{l|}{0.0003} & \multicolumn{1}{l|}{0.0706} & \multicolumn{1}{l|}{94.4} & 92.6 \\ \cline{1-7} \cline{9-15} 
\multicolumn{1}{|l|}{\multirow{5}{*}{2}} & \multicolumn{1}{l|}{Min. adj. IPW (w/ $g_0$)} & \multicolumn{1}{l|}{0.0017} & \multicolumn{1}{l|}{0.0023} & \multicolumn{1}{l|}{0.1903} & \multicolumn{1}{l|}{94.7} & \multicolumn{1}{l|}{92.7} &  & \multirow{5}{*}{9} & Min. adj. IPW (w/ $g_0$) & 0.0002 & \multicolumn{1}{l|}{0.0015} & \multicolumn{1}{l|}{0.1473} & \multicolumn{1}{l|}{95.1} & 92.7 \\ \cline{2-7} \cline{10-15} 
\multicolumn{1}{|l|}{} & \multicolumn{1}{l|}{Min. adj. IPW (w/ $g_n$)} & \multicolumn{1}{l|}{0.0023} & \multicolumn{1}{l|}{0.0007} & \multicolumn{1}{l|}{0.1910} & \multicolumn{1}{l|}{100.0} & \multicolumn{1}{l|}{100.0} &  &  & Min. adj. IPW (w/ $g_n$) & 0.0004 & \multicolumn{1}{l|}{0.0005} & \multicolumn{1}{l|}{0.1483} & \multicolumn{1}{l|}{99.7} & 100.0 \\ \cline{2-7} \cline{10-15} 
\multicolumn{1}{|l|}{} & \multicolumn{1}{l|}{Full adj. IPW} & \multicolumn{1}{l|}{0.0030} & \multicolumn{1}{l|}{0.0007} & \multicolumn{1}{l|}{0.1983} & \multicolumn{1}{l|}{100.0} & \multicolumn{1}{l|}{100.0} &  &  & Full adj. IPW & 0.0001 & \multicolumn{1}{l|}{0.0005} & \multicolumn{1}{l|}{0.1554} & \multicolumn{1}{l|}{100.0} & 100.0 \\ \cline{2-7} \cline{10-15} 
\multicolumn{1}{|l|}{} & \multicolumn{1}{l|}{Full adj. G-comp.} & \multicolumn{1}{l|}{0.0011} & \multicolumn{4}{c|}{N/A} &  &  & Full adj. G-comp. & 0.0073 & \multicolumn{4}{c|}{N/A} \\ \cline{2-7} \cline{10-15} 
\multicolumn{1}{|l|}{} & \multicolumn{1}{l|}{Full adj. TMLE} & \multicolumn{1}{l|}{0.0017} & \multicolumn{1}{l|}{0.0005} & \multicolumn{1}{l|}{0.0861} & \multicolumn{1}{l|}{95.0} & \multicolumn{1}{l|}{92.6} &  &  & Full adj. TMLE & 0.0003 & \multicolumn{1}{l|}{0.0003} & \multicolumn{1}{l|}{0.0713} & \multicolumn{1}{l|}{94.6} & 92.6 \\ \cline{1-7} \cline{9-15} 
\multicolumn{1}{|l|}{\multirow{5}{*}{3}} & \multicolumn{1}{l|}{Min. adj. IPW (w/ $g_0$)} & \multicolumn{1}{l|}{0.0017} & \multicolumn{1}{l|}{0.0023} & \multicolumn{1}{l|}{0.1898} & \multicolumn{1}{l|}{95.6} & \multicolumn{1}{l|}{92.7} &  & \multirow{5}{*}{10} & Min. adj. IPW (w/ $g_0$) & 0.0009 & \multicolumn{1}{l|}{0.0023} & \multicolumn{1}{l|}{0.1885} & \multicolumn{1}{l|}{95.5} & 92.7 \\ \cline{2-7} \cline{10-15} 
\multicolumn{1}{|l|}{} & \multicolumn{1}{l|}{Min. adj. IPW (w/ $g_n$)} & \multicolumn{1}{l|}{0.0001} & \multicolumn{1}{l|}{0.0008} & \multicolumn{1}{l|}{0.1912} & \multicolumn{1}{l|}{99.8} & \multicolumn{1}{l|}{100.0} &  &  & Min. adj. IPW (w/ $g_n$) & 0.0013 & \multicolumn{1}{l|}{0.0008} & \multicolumn{1}{l|}{0.1895} & \multicolumn{1}{l|}{99.8} & 100.0 \\ \cline{2-7} \cline{10-15} 
\multicolumn{1}{|l|}{} & \multicolumn{1}{l|}{Full adj. IPW} & \multicolumn{1}{l|}{0.0015} & \multicolumn{1}{l|}{0.0007} & \multicolumn{1}{l|}{0.1989} & \multicolumn{1}{l|}{99.9} & \multicolumn{1}{l|}{100.0} &  &  & Full adj. IPW & 0.0001 & \multicolumn{1}{l|}{0.0008} & \multicolumn{1}{l|}{0.1971} & \multicolumn{1}{l|}{100.0} & 100.0 \\ \cline{2-7} \cline{10-15} 
\multicolumn{1}{|l|}{} & \multicolumn{1}{l|}{Full adj. G-comp.} & \multicolumn{1}{l|}{0.0091} & \multicolumn{4}{c|}{N/A} &  &  & Full adj. G-comp. & 0.0131 & \multicolumn{4}{c|}{N/A} \\ \cline{2-7} \cline{10-15} 
\multicolumn{1}{|l|}{} & \multicolumn{1}{l|}{Full adj. TMLE} & \multicolumn{1}{l|}{0.0010} & \multicolumn{1}{l|}{0.0005} & \multicolumn{1}{l|}{0.0865} & \multicolumn{1}{l|}{95.3} & \multicolumn{1}{l|}{92.6} &  &  & Full adj. TMLE & 0.0011 & \multicolumn{1}{l|}{0.0005} & \multicolumn{1}{l|}{0.0871} & \multicolumn{1}{l|}{93.7} & 92.6 \\ \cline{1-7} \cline{9-15} 
\multicolumn{1}{|l|}{\multirow{5}{*}{4}} & \multicolumn{1}{l|}{Min. adj. IPW (w/ $g_0$)} & \multicolumn{1}{l|}{0.0006} & \multicolumn{1}{l|}{0.0026} & \multicolumn{1}{l|}{0.1920} & \multicolumn{1}{l|}{94.2} & \multicolumn{1}{l|}{92.7} &  & \multirow{5}{*}{11} & Min. adj. IPW (w/ $g_0$) & 0.0003 & \multicolumn{1}{l|}{0.0023} & \multicolumn{1}{l|}{0.1906} & \multicolumn{1}{l|}{94.7} & 92.7 \\ \cline{2-7} \cline{10-15} 
\multicolumn{1}{|l|}{} & \multicolumn{1}{l|}{Min. adj. IPW (w/ $g_n$)} & \multicolumn{1}{l|}{0.0009} & \multicolumn{1}{l|}{0.0008} & \multicolumn{1}{l|}{0.1930} & \multicolumn{1}{l|}{99.8} & \multicolumn{1}{l|}{100.0} &  &  & Min. adj. IPW (w/ $g_n$) & 0.0010 & \multicolumn{1}{l|}{0.0008} & \multicolumn{1}{l|}{0.1917} & \multicolumn{1}{l|}{99.8} & 100.0 \\ \cline{2-7} \cline{10-15} 
\multicolumn{1}{|l|}{} & \multicolumn{1}{l|}{Full adj. IPW} & \multicolumn{1}{l|}{0.0016} & \multicolumn{1}{l|}{0.0008} & \multicolumn{1}{l|}{0.2010} & \multicolumn{1}{l|}{99.9} & \multicolumn{1}{l|}{100.0} &  &  & Full adj. IPW & 0.0004 & \multicolumn{1}{l|}{0.0007} & \multicolumn{1}{l|}{0.1995} & \multicolumn{1}{l|}{100.0} & 100.0 \\ \cline{2-7} \cline{10-15} 
\multicolumn{1}{|l|}{} & \multicolumn{1}{l|}{Full adj. G-comp.} & \multicolumn{1}{l|}{0.0066} & \multicolumn{4}{c|}{N/A} &  &  & Full adj. G-comp. & 0.0029 & \multicolumn{4}{c|}{N/A} \\ \cline{2-7} \cline{10-15} 
\multicolumn{1}{|l|}{} & \multicolumn{1}{l|}{Full adj. TMLE} & \multicolumn{1}{l|}{0.0008} & \multicolumn{1}{l|}{0.0005} & \multicolumn{1}{l|}{0.0852} & \multicolumn{1}{l|}{94.1} & \multicolumn{1}{l|}{92.6} &  &  & Full adj. TMLE & 0.0006 & \multicolumn{1}{l|}{0.0005} & \multicolumn{1}{l|}{0.0860} & \multicolumn{1}{l|}{94.3} & 92.6 \\ \cline{1-7} \cline{9-15} 
\multicolumn{1}{|l|}{\multirow{5}{*}{5}} & \multicolumn{1}{l|}{Min. adj. IPW (w/ $g_0$)} & \multicolumn{1}{l|}{0.0008} & \multicolumn{1}{l|}{0.0024} & \multicolumn{1}{l|}{0.1887} & \multicolumn{1}{l|}{94.1} & \multicolumn{1}{l|}{92.7} &  & \multirow{5}{*}{12} & Min. adj. IPW (w/ $g_0$) & 0.0002 & \multicolumn{1}{l|}{0.0024} & \multicolumn{1}{l|}{0.1899} & \multicolumn{1}{l|}{94.9} & 92.7 \\ \cline{2-7} \cline{10-15} 
\multicolumn{1}{|l|}{} & \multicolumn{1}{l|}{Min. adj. IPW (w/ $g_n$)} & \multicolumn{1}{l|}{0.0004} & \multicolumn{1}{l|}{0.0008} & \multicolumn{1}{l|}{0.1896} & \multicolumn{1}{l|}{99.8} & \multicolumn{1}{l|}{100.0} &  &  & Min. adj. IPW (w/ $g_n$) & 0.0002 & \multicolumn{1}{l|}{0.0008} & \multicolumn{1}{l|}{0.1909} & \multicolumn{1}{l|}{100.0} & 100.0 \\ \cline{2-7} \cline{10-15} 
\multicolumn{1}{|l|}{} & \multicolumn{1}{l|}{Full adj. IPW} & \multicolumn{1}{l|}{0.0013} & \multicolumn{1}{l|}{0.0008} & \multicolumn{1}{l|}{0.1973} & \multicolumn{1}{l|}{100.0} & \multicolumn{1}{l|}{100.0} &  &  & Full adj. IPW & 0.0020 & \multicolumn{1}{l|}{0.0007} & \multicolumn{1}{l|}{0.1983} & \multicolumn{1}{l|}{100.0} & 100.0 \\ \cline{2-7} \cline{10-15} 
\multicolumn{1}{|l|}{} & \multicolumn{1}{l|}{Full adj. G-comp.} & \multicolumn{1}{l|}{0.0139} & \multicolumn{4}{c|}{N/A} &  &  & Full adj. G-comp. & 0.0074 & \multicolumn{4}{c|}{N/A} \\ \cline{2-7} \cline{10-15} 
\multicolumn{1}{|l|}{} & \multicolumn{1}{l|}{Full adj. TMLE} & \multicolumn{1}{l|}{0.0006} & \multicolumn{1}{l|}{0.0005} & \multicolumn{1}{l|}{0.0868} & \multicolumn{1}{l|}{94.5} & \multicolumn{1}{l|}{92.6} &  &  & Full adj. TMLE & 0.0002 & \multicolumn{1}{l|}{0.0005} & \multicolumn{1}{l|}{0.0866} & \multicolumn{1}{l|}{94.6} & 92.6 \\ \cline{1-7} \cline{9-15} 
\multicolumn{1}{|l|}{\multirow{5}{*}{6}} & \multicolumn{1}{l|}{Min. adj. IPW (w/ $g_0$)} & \multicolumn{1}{l|}{0.0004} & \multicolumn{1}{l|}{0.0026} & \multicolumn{1}{l|}{0.1910} & \multicolumn{1}{l|}{94.2} & \multicolumn{1}{l|}{92.7} &  & \multirow{5}{*}{13} & Min. adj. IPW (w/ $g_0$) & 0.0009 & \multicolumn{1}{l|}{0.0026} & \multicolumn{1}{l|}{0.1920} & \multicolumn{1}{l|}{93.9} & 92.7 \\ \cline{2-7} \cline{10-15} 
\multicolumn{1}{|l|}{} & \multicolumn{1}{l|}{Min. adj. IPW (w/ $g_n$)} & \multicolumn{1}{l|}{0.0012} & \multicolumn{1}{l|}{0.0008} & \multicolumn{1}{l|}{0.1914} & \multicolumn{1}{l|}{100.0} & \multicolumn{1}{l|}{100.0} &  &  & Min. adj. IPW (w/ $g_n$) & 0.0005 & \multicolumn{1}{l|}{0.0008} & \multicolumn{1}{l|}{0.1930} & \multicolumn{1}{l|}{99.8} & 100.0 \\ \cline{2-7} \cline{10-15} 
\multicolumn{1}{|l|}{} & \multicolumn{1}{l|}{Full adj. IPW} & \multicolumn{1}{l|}{0.0015} & \multicolumn{1}{l|}{0.0008} & \multicolumn{1}{l|}{0.1993} & \multicolumn{1}{l|}{100.0} & \multicolumn{1}{l|}{100.0} &  &  & Full adj. IPW & 0.0017 & \multicolumn{1}{l|}{0.0007} & \multicolumn{1}{l|}{0.2007} & \multicolumn{1}{l|}{100.0} & 100.0 \\ \cline{2-7} \cline{10-15} 
\multicolumn{1}{|l|}{} & \multicolumn{1}{l|}{Full adj. G-comp.} & \multicolumn{1}{l|}{0.0020} & \multicolumn{4}{c|}{N/A} &  &  & Full adj. G-comp. & 0.0084 & \multicolumn{4}{c|}{N/A} \\ \cline{2-7} \cline{10-15} 
\multicolumn{1}{|l|}{} & \multicolumn{1}{l|}{Full adj. TMLE} & \multicolumn{1}{l|}{0.0006} & \multicolumn{1}{l|}{0.0005} & \multicolumn{1}{l|}{0.0857} & \multicolumn{1}{l|}{94.6} & \multicolumn{1}{l|}{92.6} &  &  & Full adj. TMLE & 0.0003 & \multicolumn{1}{l|}{0.0005} & \multicolumn{1}{l|}{0.0853} & \multicolumn{1}{l|}{94.3} & 92.6 \\ \cline{1-7} \cline{9-15} 
\multicolumn{1}{|l|}{\multirow{5}{*}{7}} & \multicolumn{1}{l|}{Min. adj. IPW (w/ $g_0$)} & \multicolumn{1}{l|}{0.0012} & \multicolumn{1}{l|}{0.0015} & \multicolumn{1}{l|}{0.1469} & \multicolumn{1}{l|}{94.7} & \multicolumn{1}{l|}{92.7} &  & \multirow{5}{*}{14} & Min. adj. IPW (w/ $g_0$) & 0.0005 & \multicolumn{1}{l|}{0.0023} & \multicolumn{1}{l|}{0.1888} & \multicolumn{1}{l|}{95.1} & 92.7 \\ \cline{2-7} \cline{10-15} 
\multicolumn{1}{|l|}{} & \multicolumn{1}{l|}{Min. adj. IPW (w/ $g_n$)} & \multicolumn{1}{l|}{0.0007} & \multicolumn{1}{l|}{0.0005} & \multicolumn{1}{l|}{0.1476} & \multicolumn{1}{l|}{100.0} & \multicolumn{1}{l|}{100.0} &  &  & Min. adj. IPW (w/ $g_n$) & 0.0001 & \multicolumn{1}{l|}{0.0008} & \multicolumn{1}{l|}{0.1895} & \multicolumn{1}{l|}{99.9} & 100.0 \\ \cline{2-7} \cline{10-15} 
\multicolumn{1}{|l|}{} & \multicolumn{1}{l|}{Full adj. IPW} & \multicolumn{1}{l|}{0.0003} & \multicolumn{1}{l|}{0.0005} & \multicolumn{1}{l|}{0.1536} & \multicolumn{1}{l|}{99.9} & \multicolumn{1}{l|}{100.0} &  &  & Full adj. IPW & 0.0012 & \multicolumn{1}{l|}{0.0008} & \multicolumn{1}{l|}{0.1970} & \multicolumn{1}{l|}{100.0} & 100.0 \\ \cline{2-7} \cline{10-15} 
\multicolumn{1}{|l|}{} & \multicolumn{1}{l|}{Full adj. G-comp.} & \multicolumn{1}{l|}{0.0056} & \multicolumn{4}{c|}{N/A} &  &  & Full adj. G-comp. & 0.0122 & \multicolumn{4}{c|}{N/A} \\ \cline{2-7} \cline{10-15} 
\multicolumn{1}{|l|}{} & \multicolumn{1}{l|}{Full adj. TMLE} & \multicolumn{1}{l|}{0.0003} & \multicolumn{1}{l|}{0.0003} & \multicolumn{1}{l|}{0.0719} & \multicolumn{1}{l|}{93.5} & \multicolumn{1}{l|}{92.6} &  &  & Full adj. TMLE & 0.0010 & \multicolumn{1}{l|}{0.0005} & \multicolumn{1}{l|}{0.0869} & \multicolumn{1}{l|}{94.1} & 92.6 \\ \cline{1-7} \cline{9-15} 
 &  &  &  &  &  &  &  & \multirow{5}{*}{15} & Min. adj. IPW (w/ $g_0$) & 0.0001 & \multicolumn{1}{l|}{0.0023} & \multicolumn{1}{l|}{0.1909} & \multicolumn{1}{l|}{94.3} & 92.7 \\ \cline{10-15} 
 &  &  &  &  &  &  &  &  & Min. adj. IPW (w/ $g_n$) & 0.0001 & \multicolumn{1}{l|}{0.0007} & \multicolumn{1}{l|}{0.1917} & \multicolumn{1}{l|}{99.9} & 100.0 \\ \cline{10-15} 
 &  &  &  &  &  &  &  &  & Full adj. IPW & 0.0008 & \multicolumn{1}{l|}{0.0007} & \multicolumn{1}{l|}{0.1993} & \multicolumn{1}{l|}{99.9} & 100.0 \\ \cline{10-15} 
 &  &  &  &  &  &  &  &  & Full adj. G-comp. & 0.0036 & \multicolumn{4}{c|}{N/A} \\ \cline{10-15} 
 &  &  &  &  &  &  &  &  & Full adj. TMLE & 0.0005 & \multicolumn{1}{l|}{0.0005} & \multicolumn{1}{l|}{0.0857} & \multicolumn{1}{l|}{94.9} & 92.6 \\ \cline{9-15} 
\end{tabular}
}
\label{DGP2table}
\caption{(Caption on the following page.)}
\end{sidewaystable}

\begin{table}
  \contcaption{DGP 2: Performance (absolute bias [Abs. Bias], variance [Var.], confidence interval [C.I.] width, individual [Ind.] and simultaneous [Simult.] confidence interval coverage [Cov.]) of each of the 5 estimators for the values of 15 regimes embedded in outcome-blind simulations of the Adaptive Strategies for Preventing and Treating Lapses of Retention in HIV Care (ADAPT-R) trial data. The 5 estimators evaluated are: 1) an inverse probability weighted (IPW) estimator with weights based on the true, known probability of receiving treatment (``Min. adj IPW (w/ $g_0$)"); 2) an IPW estimator with estimated weights based on the empirical proportion of receiving treatment, which is equivalent to a TMLE or G-computation estimator where iterated conditional expectation (ICE) factors are estimated with saturated regression models non-parametric maximum likelihood estimator (``Min. adj IPW (w/ $g_n$)"); 3) an IPW estimator with estimated weights that adjust for all covariates (``Full adj. IPW"); 4) a G-computation estimator based on ICEs estimated with machine learning that adjust for all covariates (``Full adj. G-comp."); and 5) a targeted maximum likelihood estimator (TMLE) that adjusts for all covariates (``Full adj. TMLE").}
\end{table}

\subsection{Appendix D}

\begin{figure}[h]
    \centering
    \includegraphics[scale = .65]{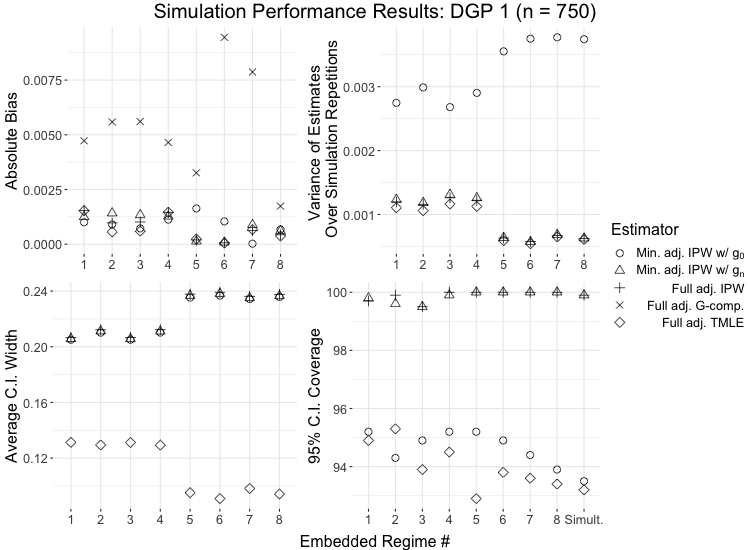}
    \caption{DGP 1 with sample size $n=$750. Performance (top left panel is absolute bias, top right panel is Monte Carlo variance (over simulation repetitions), bottom left panel is mean confidence interval [C.I.] width across simulation repetitions, and bottom right panel is 95\% C.I. coverage) of candidate estimators of the value of each of the 8 embedded regimes within the simple sequential multiple assignment randomized trial (SMART) generated from DGP 1. The 5 estimators evaluated are: 1) an inverse probability weighted (IPW) estimator with weights based on the true, known probability of receiving treatment given the initial treatment and lapse response (``Min. adj IPW (w/ $g_0$)"); 2) an IPW estimator with estimated weights based on the empirical proportion of receiving treatment given the initial treatment and lapse response, which is equivalent to a TMLE or G-computation estimator where iterated conditional expectation (ICE) factors are estimated with saturated regression models (``Min. adj IPW (w/ $g_n$)"); 3) an IPW estimator with estimated weights that adjust for all covariates (``Full adj. IPW"); 4) a G-computation estimator based on ICEs estimated with machine learning that adjust for all covariates (``Full adj. G-comp."); and 5) a targeted maximum likelihood estimator (TMLE) that adjusts for all covariates (``Full adj. TMLE"). Both individual and simultaneous C.I. coverage is shown under the regime numbers 1-8 and ``Simult.," respectively.}
    \label{DGP1fig_750}
\end{figure}

\begin{figure}[h]
    \centering
    \includegraphics[scale = .65]{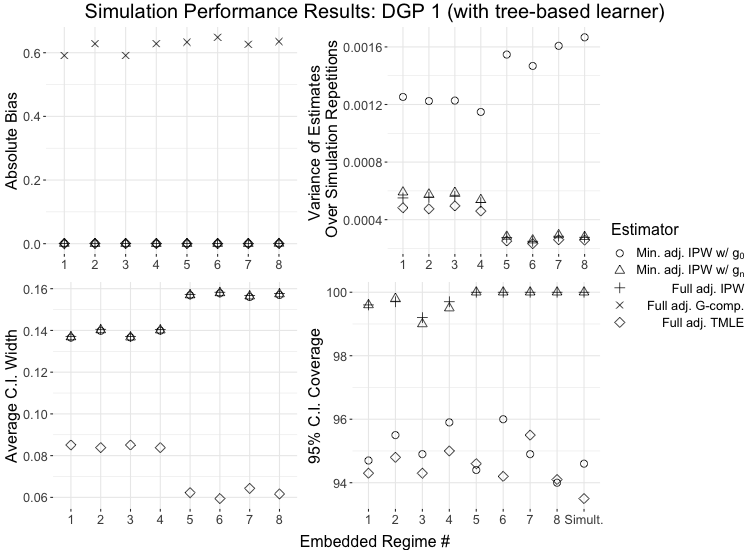}
    \caption{DGP 1 with SuperLearner library that includes a tree-based algorithm in the SuperLearner. Performance (top left panel is absolute bias, top right panel is Monte Carlo variance (over simulation repetitions), bottom left panel is mean confidence interval [C.I.] width across simulation repetitions, and bottom right panel is 95\% C.I. coverage) of candidate estimators of the value of each of the 8 embedded regimes within the simple sequential multiple assignment randomized trial (SMART) generated from DGP 1. The 5 estimators evaluated are: 1) an inverse probability weighted (IPW) estimator with weights based on the true, known probability of receiving treatment given the initial treatment and lapse response (``Min. adj IPW (w/ $g_0$)"); 2) an IPW estimator with estimated weights based on the empirical proportion of receiving treatment given the initial treatment and lapse response, which is equivalent to a TMLE or G-computation estimator where iterated conditional expectation (ICE) factors are estimated with saturated regression models (``Min. adj IPW (w/ $g_n$)"); 3) an IPW estimator with estimated weights that adjust for all covariates (``Full adj. IPW"); 4) a G-computation estimator based on ICEs estimated with machine learning that adjust for all covariates (``Full adj. G-comp."); and 5) a targeted maximum likelihood estimator (TMLE) that adjusts for all covariates (``Full adj. TMLE"). Both individual and simultaneous C.I. coverage is shown under the regime numbers 1-8 and ``Simult.," respectively.}
    \label{DGP1fig_tree}
\end{figure}

\begin{figure}[h]
    \centering
    \includegraphics[scale = .65]{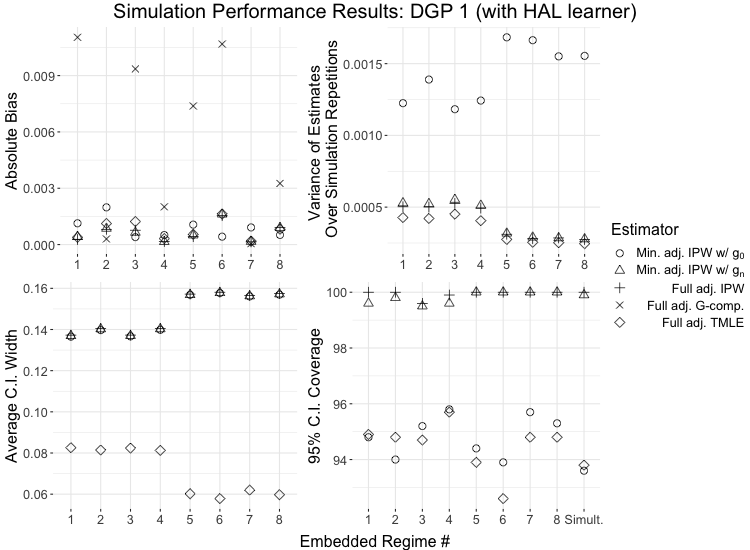}
    \caption{ DGP 1 with SuperLearner library that includes highly adaptive lasso (HAL) in the SuperLearner. Performance (top left panel is absolute bias, top right panel is Monte Carlo variance (over simulation repetitions), bottom left panel is mean confidence interval [C.I.] width across simulation repetitions, and bottom right panel is 95\% C.I. coverage) of candidate estimators of the value of each of the 8 embedded regimes within the simple sequential multiple assignment randomized trial (SMART) generated from DGP 1. The 5 estimators evaluated are: 1) an inverse probability weighted (IPW) estimator with weights based on the true, known probability of receiving treatment given the initial treatment and lapse response (``Min. adj IPW (w/ $g_0$)"); 2) an IPW estimator with estimated weights based on the empirical proportion of receiving treatment given the initial treatment and lapse response, which is equivalent to a TMLE or G-computation estimator where iterated conditional expectation (ICE) factors are estimated with saturated regression models (``Min. adj IPW (w/ $g_n$)"); 3) an IPW estimator with estimated weights that adjust for all covariates (``Full adj. IPW"); 4) a G-computation estimator based on ICEs estimated with machine learning that adjust for all covariates (``Full adj. G-comp."); and 5) a targeted maximum likelihood estimator (TMLE) that adjusts for all covariates (``Full adj. TMLE"). Both individual and simultaneous C.I. coverage is shown under the regime numbers 1-8 and ``Simult.," respectively.}
    \label{DGP1fig_HAL}
\end{figure}

In this section we present figures for the performance results for alternative simulation configurations for DGP 1, including: 1) the same estimator configurations and comparisons as the main text but with a smaller sample size ($n=$750; Figure \ref{DGP1fig_750}); 2) the same estimator comparisons with estimators that include a SuperLearner library with a tree-based algorithm (Figure \ref{DGP1fig_tree}); 3) the same estimator comparisons with estimators that include a SuperLearner library using HAL \citep{benkeser2016highly, ertefaie2020nonparametric} (Figure \ref{DGP1fig_HAL}).


\subsection{Appendix E}

In this section we present tables corresponding to the ADAPT-R analysis described in the main text. Table \ref{adapt_results_tab} shows the TMLE point estimates (with individual and simultaneous confidence intervals) of the probability of viral suppression for each of the 15 embedded regimes, in addition to the number of participants who followed each regime in the study. Table \ref{adapt_results_contrast_tab} shows estimates of ADAPT-R's pre-specified contrasts (using TMLE and three versions of IPW) with confidence intervals, in addition to the relative confidence interval widths between the non-TMLE estimators and TMLE to illustrate TMLE's variance gains.

\begin{table}[]
\begin{tabular}{|l|l|l|l|l|}
\hline
Embedded Regime \# & Num. Follow & Point Estimate & {[}95\% Simult. CI{]} & {[}95\% Ind. CI{]} \\ \hline
1 & 489 & 0.7475 & {[}0.6843, 0.8107{]} & {[}0.7044, 0.7906{]} \\ \hline
2 & 237 & 0.7834 & {[}0.7084, 0.8583{]} & {[}0.7323, 0.8345{]} \\ \hline
3 & 291 & 0.8045 & {[}0.7433, 0.8657{]} & {[}0.7628, 0.8462{]} \\ \hline
4 & 486 & 0.7976 & {[}0.7352, 0.8600{]} & {[}0.7550, 0.8401{]} \\ \hline
5 & 252 & 0.7779 & {[}0.7088, 0.8470{]} & {[}0.7308, 0.8250{]} \\ \hline
6 & 282 & 0.8135 & {[}0.7500, 0.8770{]} & {[}0.7702, 0.8568{]} \\ \hline
7 & 489 & 0.7912 & {[}0.7308, 0.8516{]} & {[}0.7500, 0.8324{]} \\ \hline
8 & 244 & 0.8280 & {[}0.7621, 0.8939{]} & {[}0.7831, 0.8729{]} \\ \hline
9 & 298 & 0.8300 & {[}0.7717, 0.8883{]} & {[}0.7902, 0.8697{]} \\ \hline
10 & 268 & 0.7643 & {[}0.6919, 0.8366{]} & {[}0.7149, 0.8136{]} \\ \hline
11 & 341 & 0.7287 & {[}0.6623, 0.7952{]} & {[}0.6834, 0.7740{]} \\ \hline
12 & 283 & 0.7583 & {[}0.6918, 0.8248{]} & {[}0.7130, 0.8036{]} \\ \hline
13 & 332 & 0.7372 & {[}0.6682, 0.8062{]} & {[}0.6902, 0.7842{]} \\ \hline
14 & 275 & 0.8098 & {[}0.7472, 0.8724{]} & {[}0.7672, 0.8525{]} \\ \hline
15 & 348 & 0.7530 & {[}0.6889, 0.8171{]} & {[}0.7093, 0.7967{]} \\ \hline
\end{tabular}
\caption{Estimated values of the Adaptive Strategies for Preventing and Treating Lapses of Retention in HIV Care (ADAPT-R) trial's 15 embedded regimes listed in Table 1 in the main text. The second column ``Num. Follow" lists the number of patients in ADAPT-R who contributed to that regime. Targeted Maximum Likelihood Estimation (TMLE) was used to calculate point estimates; simultaneous (``Simult.") and individual (``Ind.") confidence intervals (CIs) were calculated using the estimated efficient influence curve.}
\label{adapt_results_tab}
\end{table}

\begin{table}[]
\begin{tabular}{|l|l|l|l|}
\hline
Comparator \\ embedded regime \# & Estimator & Difference Estimate {[}95\% CI{]} & Relative CI Width \\ \hline
\multirow{4}{*}{5} & Min. adj. IPW (w/ $g_0$) & -0.1099 {[}-0.2343, 0.0144{]} & 1.96 \\ \cline{2-4} 
 & Min. adj. IPW (w/ $g_n$) & 0.0210 {[}-0.1156, 0.1577{]} & 2.15 \\ \cline{2-4} 
 & Full adj. IPW & 0.0338 {[}-0.1036, 0.1712{]} & 2.16 \\ \cline{2-4} 
 & Full adj. TMLE & 0.0304 {[}-0.0331, 0.0939{]} & -- \\ \hline
\multirow{4}{*}{6} & Min. adj. IPW (w/ $g_0$) & 0.0248 {[}-0.1063, 0.1559{]} & 2.16 \\ \cline{2-4} 
 & Min. adj. IPW (w/ $g_n$) & 0.0595 {[}-0.0741, 0.1931{]} & 2.20 \\ \cline{2-4} 
 & Full adj. IPW & 0.0649 {[}-0.0661, 0.1960{]} & 2.16 \\ \cline{2-4} 
 & Full adj. TMLE & 0.0660 {[}0.0053, 0.1267{]} & -- \\ \hline
\multirow{4}{*}{8} & Min. adj. IPW (w/ $g_0$) & -0.0780 {[}-0.205, 0.0490{]} & 2.05 \\ \cline{2-4} 
 & Min. adj. IPW (w/ $g_n$) & 0.0747 {[}-0.0675, 0.2169{]} & 2.30 \\ \cline{2-4} 
 & Full adj. IPW & 0.1023 {[}-0.0464, 0.2510{]} & 2.40 \\ \cline{2-4} 
 & Full adj. TMLE & 0.0805 {[}0.0185, 0.1424{]} & -- \\ \hline
\multirow{4}{*}{9} & Min. adj. IPW (w/ $g_0$) & 0.0887 {[}-0.0473, 0.2246{]} & 2.33 \\ \cline{2-4} 
 & Min. adj. IPW (w/ $g_n$) & 0.0797 {[}-0.0523, 0.2118{]} & 2.26 \\ \cline{2-4} 
 & Full adj. IPW & 0.1037 {[}-0.0292, 0.2367{]} & 2.28 \\ \cline{2-4} 
 & Full adj. TMLE & 0.0825 {[}0.0241, 0.1408{]} & -- \\ \hline
\multirow{4}{*}{11} & Min. adj. IPW (w/ $g_0$) & 0.1064 {[}-0.0290, 0.2418{]} & 2.18 \\ \cline{2-4} 
 & Min. adj. IPW (w/ $g_n$) & -0.0227 {[}-0.1432, 0.0979{]} & 1.94 \\ \cline{2-4} 
 & Full adj. IPW & -0.0103 {[}-0.1297, 0.1092{]} & 1.92 \\ \cline{2-4} 
 & Full adj. TMLE & -0.0188 {[}-0.0810, 0.0434{]} & -- \\ \hline
\end{tabular}
\caption{Table 6. Pre-specified contrast analysis of the Adaptive Strategies for Preventing and Treating Lapses of Retention in HIV Care (ADAPT-R) study. Estimates of the difference in probability of viral suppression for the following pre-specified rules compared to standard of care (SOC) throughout: 1) short message service (SMS) with continuation if no lapse and addition of conditional cash transfer (CCT) if a lapse occurred (embedded regime number 5); 2) CCT with continuation if no lapse and addition of SMS if lapse occurred (embedded regime number 6); 3) SMS with continuation if no lapse and replacement with navigator if lapse occurred (regime number 8); 4) CCT with continuation of no lapse and replacement of navigator if lapse occurred (regime number 9); and 5) initial CCT, with SOC outreach if a lapse occurred and disconuation if no lapse occurred (regime number 11). Difference estimates (with influence curve-based individual confidence intervals [CIs]) were generated with: 1) an inverse probability weighted (IPW) estimator with weights based on the true, known probability of receiving treatment given the initial treatment and lapse response (``Min. adj IPW w/ $g_0$"); 2) an IPW estimator with estimated weights based on the empirical proportion of receiving treatment given the initial treatment and lapse response (``Min. adj IPW w/ $g_n$"); 3) an IPW estimator with estimated weights that adjust for all covariates (``Full adj. IPW"); and 4) a targeted maximum likelihood estimator (TMLE) that adjusts for all covariates (``Full adj. TMLE"). The last column denotes the relative CI  width between the estimator in that row and the ``Full adj. TMLE."}
\label{adapt_results_contrast_tab}
\end{table}

\end{document}